\documentclass{osa-article}

\journal{oe}
\usepackage{color,bm,graphicx,subfigure,ulem,soul}

\begin{document}
\title{Oceanic non-Kolmogorov optical turbulence and spherical wave propagation}

\author{Jin-Ren Yao,\authormark{1,3} Han-Tao Wang,\authormark{1} Hua-Jun Zhang,\authormark{1} Jian-Dong Cai,\authormark{1} Ming-Yuan Ren,\authormark{1} Yu Zhang,\authormark{1,*} and Olga Korotkova,\authormark{2}}

\address{\authormark{1}School of Physics, Harbin Institute of Technology, Harbin, 150001, China\\
	\authormark{2}Department of Physics, University of Miami, Coral Gables, FL 33146, USA\\
	\authormark{3}jinry@yahoo.com
	}

\email{\authormark{*}Corresponding author: zhangyuhitphy@163.com} 
{
\fontsize{8pt}{0pt}
$\\$
\textcolor{magenta}{\textit{*This paper has been published in Optics Express https://doi.org/10.1364/OE.409498. This PDF you are reading is a further modified version where we adjusted few words.}}
}
\begin{abstract}
Light propagation in turbulent media is conventionally studied with the help of the spatio-temporal power spectra of the refractive index fluctuations. In particular, for natural water turbulence several models for the spatial power spectra have been developed based on the classic, Kolmogorov postulates. However, as currently widely accepted, non-Kolmogorov turbulent regime is also common in the stratified flow fields, as suggested by recent  developments in atmospheric optics. Until now all the models developed for the non-Kolmogorov optical turbulence were pertinent to atmospheric research and, hence, involved only one advected scalar, e.g., temperature. We generalize the oceanic spatial power spectrum, based on two advected scalars, temperature and salinity concentration, to the non-Kolmogorov turbulence regime, with the help of the so-called "Upper-Bound Limitation" and by adopting the concept of spectral correlation of two advected scalars. The proposed power spectrum can handle general non-Kolmogorov, anisotropic  turbulence but reduces to Kolmogorov, isotropic case if the power law exponents of temperature and salinity are set to 11/3 and anisotropy coefficient is set to unity. To show the application of the new spectrum, we derive the expression for the second-order mutual coherence function of the spherical wave and examine its coherence radius (in both  scalar and vector forms) to characterize the turbulent disturbance. Our numerical calculations show that the statistics of the spherical wave vary substantially with temperature and salinity non-Kolmogorov power law exponents and temperature-salinity spectral correlation coefficient. The introduced spectrum is envisioned to become of significance for theoretical analysis and experimental measurements of non-classic natural water double-diffusion turbulent regimes. 
\end{abstract}

\section{Introduction}
\setulcolor{red}
The stable temperature stratification and the stable salinity stratification in marine environment may be disturbed by the velocity field, which results in the inhomogeneous spatio-temporal distribution of temperature and salinity, and hence leads to the spatio-temporal fluctuations of refractive-index. We name such refractive-index fluctuations as oceanic optical turbulence \cite{korotkova2019light}. Besides, the oceanic optical turbulence could also be driven by other factors like winds, gravity and vertical currents.
The Oceanic Turbulence Optical Power Spectrum (OTOPS) being the Fourier transform of the spatial covariance function of the refractive index provides an essential tool for characterizing the spatial statistics of any order for stationary light fields propagating through the natural waters. Within the last two decades, the oceanic power spectrum model developed in \cite{Nikishov_2000} based on Kolmogorov turbulence theory resulted, with the help of the Rytov and the extended Huygens-Fresnel methods, in a number of theoretical predictions relating to light interaction with turbulent waters. In particular, evolution of the spectral density \cite{Lu_2006}, the spectral shifts \cite{shchepakina2011spectral}, the polarimetric \cite{korotkova2011effect} and coherence \cite{farwell2012intensity} changes and propagation of several other 2nd-order and 4th-order statistics \cite{korotkova2012light,baykal2016scintillation,ata2014structure,lu2014wave} have been revealed.
The theory has also benefited a number of underwater applications, such as the oceanic Light Detection and Ranging (Lidar) \cite{korotkova2018enhanced} systems, underwater optical communications \cite{baykal2018bit,yi2015underwater,cui2019scintillation}, and underwater imaging \cite{hou2009simple}.

Since the oceanic optical turbulence is governed by two scalar fields, temperature and salinity concentration,  the OTOPS is approximately expressed as a linear combination of temperature power spectrum, salinity power spectrum and their co-spectrum \cite{Nikishov_2000}. Therefore OTOPS contains many parameters, such as the Kolmogorov scale $\eta$, the Prandtl number $Pr$, the Schmidt number $Sc$, as well as the dissipation rates of temperature, salinity, and kinetic energy, $\chi_T$, $\chi_S$, and $\varepsilon$, respectively, substantially complicating the predictions for the light - oceanic turbulence interactions.

The OTOPS model of \cite{Nikishov_2000} and its derivatives \cite{Yao_17,Elamassie:17}     
were all based on the first of the four models (called below H1) for a single-scalar turbulent advection developed by Hill \cite{Hill1978}. An alternative model for the Kolmogorov oceanic optical turbulence has been recently obtained in \cite{Yi_18,Li_19,Yao_19,KOROTKOVA2020_1} by numerically fitting model 4 of the Hill's paper (called below H4) \cite{Hill1978}. The H4-based models are more precise than the H1-based models in high spatial  frequency region, and, hence, have advantages in oceanic cases with the wide-ranged Prandtl/Schmidt numbers \cite{Muschinski:15,Yao_19}. 
\textit{All the aforementioned OTOPS models are based on the Kolmogorov theory having a constant power law $-11/3$, and the co-spectra in these models are obtained by analogy with a single scalar (temperature or salinity) spectrum.}

Kolmogorov theory relies on several assumptions including the homogeneous and isotropic nature of turbulent eddies. Such regime is clearly not universal, since it is not being able to account for several anomalous phenomena such as ramp-cliff signature and unusual scaling exponent (e. g.  \cite{sreenivasan2019turbulent}). Over the past 30 years, several experiments have revealed the presence of non-classic atmospheric optical turbulence \cite{belen1997experimental,stribling1995optical,kyrazis1994measurement,otten1999high,zilberman2005lidar,Muschinski:17}. The power spectrum model of the non-Kolmogorov turbulence advected by a single scalar and light interaction with such turbulence have been widely discussed in atmospheric optics literature  \cite{Italo2007,toselli2008,Wu:10,Shchepakina:10,SchepakinaKor:10,zilberman2008lidar,sim1,sim2,Dario}. \textit{However, it is our understanding that a comprehensive non-Kolmogorov model for oceanic waters does not exist}.

Non-Kolmogorov phenomena, as a result of inadequate rate of energy cascade, are common in underdeveloped or vertically suppressed atmospheric turbulence, and do appear in stratified marine environment. In two oceanic experiments by Ichiye \cite{ichiye1972power} and by Pochapsky and Malone \cite{Pochapsky} the non-Kolmogorov fluctuations of temperature and salinity have been observed. In the Ichiye's measurement, the power law of temperature and salinity were between $-11/3$ and $-5$, which was interpreted as the result of oceanic stratification. In Pochapsky and Malone measurement, a $-4$ power law was obtained \cite{Pochapsky}.

On considering the results of these oceanic turbulence measurements and the practical need for light propagation predictions made in various oceanic turbulence regimes, \textit{we set the aim for this paper to develop an OTOPS that extends the model suggested in \cite{Yao_19,KOROTKOVA2020_1} to non-Kolmogorov regime.} This requires (I) developing the non-Kolmogorov temperature/salinity spectrum which is applicable for the marine environment with the wide-ranged Prandtl/Schmidt numbers, and (II) deriving the temperature-salinity co-spectrum which can not be directly obtained by analogy with a single-scalar spectrum, since the power law exponents of the two advected scalars can be generally different.

The paper is organized as follows: using a non-Kolmogorov structure function, we 
derive the non-Kolmogorov temperature and salinity spectra based on the H4-based model (Section 2.1); 
using the Upper-Bound limitation, we develop a temperature-salinity co-spectrum (Section 2.2); 
on combining the results for the temperature spectrum, the salinity spectrum and the co-spectrum,
we introduce a non-Kolmogorov OTOPS (NK-OTOPS) model (Section 3); we apply the NK-OTOPS model for the analysis of the spherical wave propagation (Section 4); and we summarize the obtained results (Section 5).

\section{Temperature/salinity spectra and their co-spectrum in ocean} 
The OTOPS is composed of temperature spectrum, salinity spectrum, and temperature-salinity co-spectrum. 
In this section, we will derive the non-Kolmogorov temperature/salinity spectra (Section 2.1) and the temperature-salinity co-spectrum (Section 2.2).

\subsection{Non-Kolmogorov temperature/salinity spectra}
We begin by recalling the H4-based temperature/salinity spectrum that has been developed for Kolmogorov case in \cite{Yao_19}. By comparing its structure function with the Kolmogorov structure function, we will first obtain its structure constant $C_i^2$ and its inner scale $l_{i0}$. Then, the H4-based spectrum will be modified into a non-Kolmogorov spectrum.

\subsubsection{A. H4-based temperature/salinity spectrum}
Here the H4-based temperature/salinity spectrum \cite{Yao_19} is re-organized as
\begin{equation}
	{\Phi _i}(\kappa ) = {C_k}C_i^2{\kappa ^{ - 11/3}}{g_i}(\kappa \eta ),\ {\rm{  with }}\ i \in \left\{ {T,S} \right\},
	\label{eq1}
\end{equation}
where 
$\kappa$ is the wavenumber$\left[\rm{m}^{-1}\right]$; $C_i^2$ is the structure constant (dimensionless); $C_k$ is a dimensionless constant given by
\begin{equation}
{C_k}C_i^2 = \frac{{\beta {\varepsilon ^{ - 1/3}}{\chi _i}}}{{4\pi }},
\label{eq2}
\end{equation}
$\beta$ is the Obukhov-Corrsin constant (non-dimensional); $\varepsilon$ is the dissipation rate of kinetic energy $[\rm{m}^2\rm{s}^{-3}]$; $\chi_i$ is the ensemble-averaged variance dissipation rate of temperature or salinity ($i \in \left\{{T,S}\right\}$) with unit $\rm{K^2s^{-1}}$ or $\rm{g^2s^{-1}}$;
 the non-dimensional function $g_i(x)$ is 
\begin{equation}
{g_i}(x) = \sum\limits_{j = 0}^2 {{a_j}{x^{{b_j}}}} \exp \left( { - 174.90{x^2}{c_i}^{0.96}} \right),
\label{eq3}
\end{equation}
with
\begin{equation}
\left\{ {{a_j}} \right\} = \left\{ {1,21.61{c_i}^{0.02}, - 18.18{c_i}^{0.04}} \right\},
\label{eq4}
\end{equation}
\begin{equation}
\left\{ {{b_j}} \right\} = \left\{ {0,0.61,0.55} \right\},
\label{eq5}
\end{equation}
\begin{equation}
{c_i} = {a^{4/3}}\beta{\Pr}_i^{-1},
\label{eq6}
\end{equation}
where $Pr_T$ and $Pr_S$ are the temperature Prandtl number and salinity Schmidt number, respectively, $a$ is a constant and generally equals 0.072, and $\beta$ is the Obukhov-Corrsin constant being approximately to 0.72 \cite{Hill1978}. 

\subsubsection{B. Structure constant $C_i^2$ and inner scale $l_{i0}$}
Structure constant $C_i^2$ and inner scale $l_{i0}$ are the key parameters in the turbulence structure function, and they will be obtained by comparing the corresponding structure function in the Kolmogorov case.

The structure function of Eq.(1) is
\begin{align}
\nonumber{D_i}(R) &= 8\pi \int_0^\infty  {{\kappa ^2}} {\Phi _i}(\kappa )\left( {1 - \frac{{\sin \kappa R}}{{\kappa R}}} \right)d\kappa\\
\nonumber&= \beta {\varepsilon ^{ - 1/3}}{\chi _i}{\eta ^{2/3}}\sum\limits_{j = 0}^2 {{a_j}\left\{ {{{\left( {174.90{c_i}^{0.96}} \right)}^{\frac{1}{3} - \frac{{{b_j}}}{2}}}\Gamma \left( { - \frac{1}{3} + \frac{{{b_j}}}{2}} \right)} \right.}\\
& \ \ \left. {\left[ {1 - {}_1{F_1}\left( { - \frac{1}{3} + \frac{{{b_j}}}{2},\frac{3}{2}, - \frac{{{R^2}}}{{4 \times 174.90{c_i}^{0.96}{\eta ^2}}}} \right)} \right]} \right\},
\label{eq7}
\end{align}
where $\Gamma(\cdot)$ is a Gamma function, and $_1F_1(\cdot,\cdot,\cdot)$ is a generalized hyper-geometric function.
For Kolmogorov turbulence advected by a single scalar (temperature or salinity), the structure function is (Chapter 3 of \cite{Phillips2005Laser}):
\begin{equation}
{D_i}(R) = \left\{ {\begin{array}{*{20}{c}}
	{C_i^2{l_{i0}}^{ - 4/3}{R^2}}&{R \ll {l_{i0}},}\\
	{C_i^2{R^{2/3}}}&{R \gg {l_{i0}},}
	\end{array}} \right.\ {\rm{  with }}\ i \in \left\{ {T,S} \right\}.
\label{eq8}
\end{equation}
By comparing Eqs. (\ref{eq7}) and (\ref{eq8}), and in combining with Eq. (\ref{eq2}), we get the dimensionless constant
\begin{equation}
{C_k} = \frac{{ - \Gamma \left( {11/6} \right){2^{2/3}}}}{{4\pi \Gamma \left( { - 1/3} \right)\Gamma \left( {3/2} \right)}} \approx 0.033,
\label{eq9}
\end{equation}
\textit{the structure constant}
\begin{equation}
C_i^2 =  - \beta {\varepsilon ^{ - 1/3}}{\chi _i}\Gamma \left( { - 1/3} \right)\frac{{\Gamma \left( {3/2} \right)}}{{\Gamma \left( {11/6} \right)}}{2^{ -2/3}},
\label{eq10}
\end{equation}
and \textit{the inner scale}
\begin{equation}
{l_{i0}} = \eta/{T}\left( {{{c }_i}} \right) ,
\label{eq11}
\end{equation}
where
\begin{equation}
{T}\left( {{{c }_i}} \right) = {\left\{ {\pi {C_k}\sum\limits_{j = 0}^2 {{a_j}\left[ {{{\left( {174.90{c_i}^{0.96}} \right)}^{ - \frac{{{b_j}}}{2} - \frac{2}{3}}}\left( {\frac{{3{b_j} - 2}}{9}} \right)\Gamma \left( {\frac{{3{b_j} - 2}}{6}} \right)} \right]} } \right\}^{{\rm{3/4}}}},
\label{eq12}
\end{equation}
with $a_j$, $b_j$ and $c_i$ defined in Eqs. (\ref{eq4})-(\ref{eq6}), respectively.

\subsubsection{C. Non-Kolmogorov case}
Now, we modify Eq. (1) to a non-Kolmogorov spectrum.
Following the modification in atmospheric optics \cite{Italo2007,toselli2008,xue2011}, we add two adaptive functions $A\left(\alpha _i\right)$ and $h\left(\alpha_i, c_i\right)$ to Eq. (\ref{eq1}),
\begin{equation}
{\Phi _i}(\kappa ,{\alpha _i}) =A\left( {{\alpha _i}} \right) C_i^{\rm{2}} {\kappa ^{ - {\alpha _i}}}{g_i}\left( {\kappa \eta '} \right),
\label{eq13}
\end{equation}
with
\begin{equation}
\eta ' = \frac{\eta }{{h\left( {{\alpha _i,c_i}} \right)}},
\label{eq14}
\end{equation}
where 
$A\left(\alpha_i\right)$ is a variable factor similar to the `$A\left(\alpha\right)$' in \cite{Italo2007},
$h\left(\alpha_i , c_i\right)$ is a scaling function similar to the `$c\left(\alpha\right)$' in \cite{Italo2007}, it adjusts the location of viscous range on $\kappa$-axis.
\textit{Expressions of $A\left(\alpha_i\right)$ and $h\left(\alpha_i , c_i\right)$ are derived as follows}.

The structure function of Eq.(13) is
\begin{equation}
\begin{split}
{D_i}(R,\alpha_i)&= 8\pi \int_0^\infty  {{\kappa ^2}} {\Phi _i}(\kappa ,{\alpha _i})\left( {1 - \frac{{\sin \kappa R}}{{\kappa R}}} \right)d\kappa\\
&= 4\pi C_i^{\rm{2}}A\left( {{\alpha _i}} \right){ {\eta '} ^{\alpha  - 3}}\sum\limits_{j = 0}^2 {{a_j}\left\{ {{{\left( {174.90{c_i}^{0.96}} \right)}^{\frac{{ - 3 - {b_j} + {\alpha _i}}}{2}}}\Gamma \left( {\frac{{3 + {b_j} - {\alpha _i}}}{2}} \right)} \right.}\\
&\quad\left.\left[{1{ - _1}{F_1}\left( {\frac{{3 + {b_j} - {\alpha _i}}}{2},\frac{3}{2}, - \frac{{{R^2}}}{{4\left( {174.90{c_i}^{0.96}} \right)\eta {'^2}}}} \right)} \right]\right\},
\label{eq15}
\end{split}
\end{equation}
and the non-Kolmogorov structure function \cite{Italo2007,toselli2008} is
\begin{equation}
{D_i}(R,\alpha_i) = \left\{ {\begin{array}{*{20}{c}}
	{C_i^2{l^{\alpha_i  - 5}_{i0}}{R^2}}&{R \ll {l_{i0}},}\\
	{C_i^2{R^{\alpha_i  - {\rm{3}}}}}&{R \gg {l_{i0}},}
	\end{array}} \right.
\label{eq16}
\end{equation}
where $C^2_i$ and $l_{i0}$ have been obtained in Eqs. (\ref{eq10}) and (\ref{eq11}), respectively.
By comparing Eq. (\ref{eq15}) with Eq. (\ref{eq16}), we have
\begin{equation}
A\left( {{\alpha _i}} \right) = \frac{{\Gamma \left( {{\alpha _i} - 1} \right)}}{{4{\pi ^2}}}\cos \left( {\frac{{\pi {\alpha _i}}}{2}} \right),
\label{eq17}
\end{equation}
and
\begin{equation}
h\left( {{\alpha _i}},{c_i} \right) = G\left( {{\alpha _i},{c_i}} \right){T_i}\left( {{c_i}} \right),
\label{eq18}
\end{equation}
with
\begin{equation}
G\left( {{\alpha _i},{{c}_i}} \right) = {\left[ {\pi A\left( {{\alpha _i}} \right)\sum\limits_{j = 0}^2 {{a_j}{{\left( {174.90{c_i}^{0.96}} \right)}^{\frac{{ - 5 - {b_j} + {\alpha _i}}}{2}}}\left( {\frac{{3 + {b_j} - {\alpha _i}}}{3}} \right)\Gamma \left( {\frac{{3 + {b_j} - {\alpha _i}}}{2}} \right)} } \right]^{\frac{1}{{{\alpha _i} - 5}}}},
\label{eq19}
\end{equation}
$T_i\left(c_i\right)$ has been given in Eq. (\ref{eq12}).
When $\alpha _i = 11/3$, we have $A\left( 11/3 \right) = C_k \approx 0.033$ and $h\left( {11/3},{c_i} \right) = 1$. Hence, the non-Kolmogorov spectrum Eq. (\ref{eq13}) can degenerate into the traditional Kolmogorov model Eq. (\ref{eq1}).

\begin{figure*}
	\centering
	\includegraphics[width=0.6\linewidth]{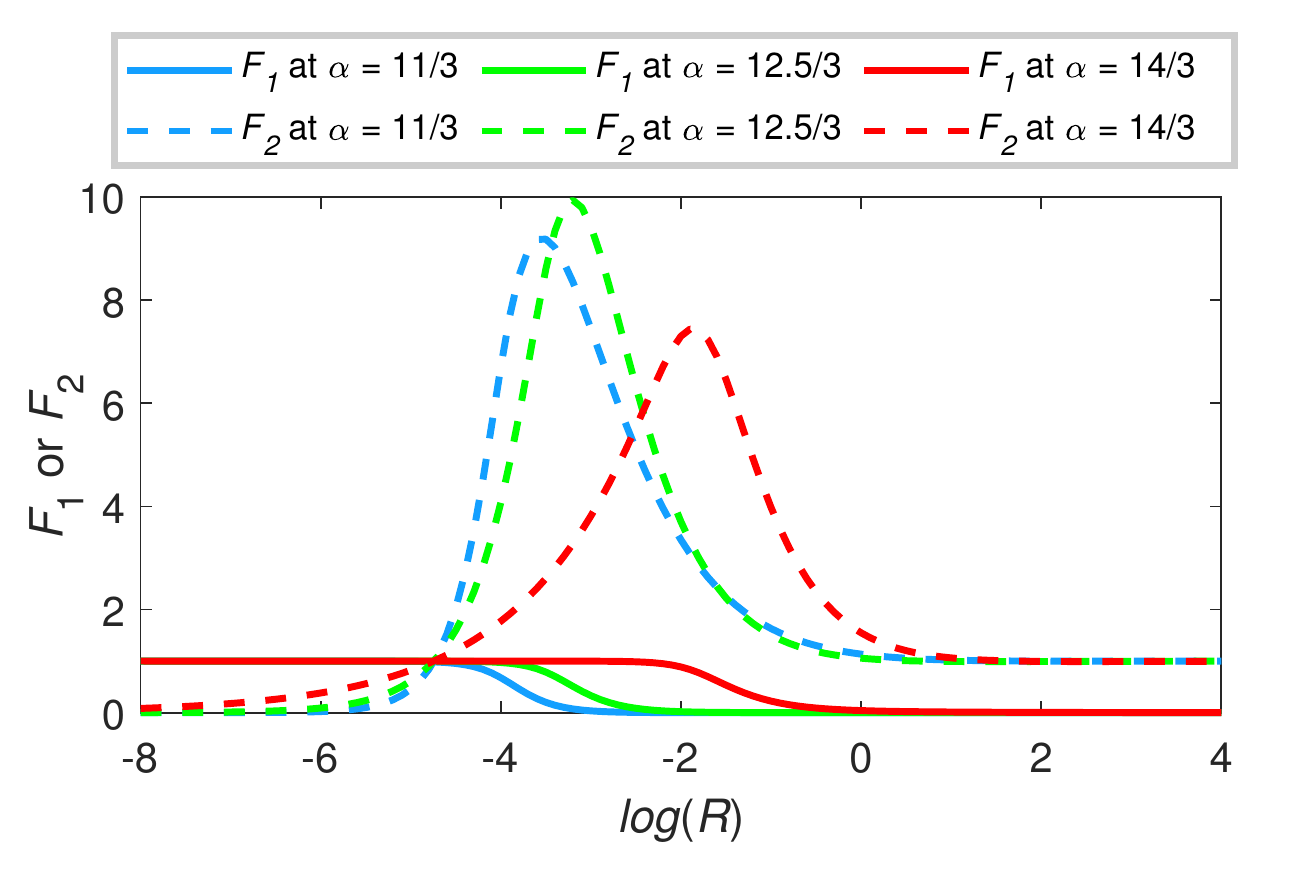}
	\caption{Functions $F_1(R)$ and $F_2(R)$ in Eq. (\ref{eq20}). The solid curves represent $F_1$, and the dashed curves represent $F_2$. $R$ has a unit [m], and the functions $F_1(R)$, $F_2(R)$ are dimensionless.}
	\label{fig:1}
\end{figure*}

To show the consistency between the proposed non-Kolmogorov spectrum Eq. (\ref{eq13}) and the non-Kolmogorov structure function Eq. (\ref{eq16}), we plot the following two functions in Fig. \ref{fig:1},
\begin{equation}
\left\{ {\begin{array}{*{20}{c}}
	{{F_1}\left( R \right) = {{\left( {C_i^2{l_{i0}}^{{\alpha _i} - 5}{R^2}} \right)}^{ - 1}}8\pi \int_0^\infty  {{\kappa ^2}} {\Phi _i}(\kappa )\left( {1 - \frac{{\sin \kappa R}}{{\kappa R}}} \right)d\kappa ,}\\
	{{F_2}\left( R \right) = {{\left( {C_i^2{R^{{\alpha _i} - 3}}} \right)}^{ - 1}}8\pi \int_0^\infty  {{\kappa ^2}} {\Phi _i}(\kappa )\left( {1 - \frac{{\sin \kappa R}}{{\kappa R}}} \right)d\kappa .}
	\end{array}} \right.
\label{eq20}
\end{equation}
It shows ${F_1}\left( {R \to 0} \right) = 1$ and ${F_2}\left( {R \to \infty } \right) = 1$, which indicates that the modified non-Kolmogorov spectrum Eq.(13) agrees well with the asymptotic formula Eq. (\ref{eq16}).

\textit{Equation (13) together with Eqs. (14), (17) and (18) constitute the main results of this section. They give the non-Kolmogorov spectrum of oceanic temperature/salinity turbulence, and the proposed spectrum agrees well with the widely accepted asymptotic structure function.} It must be noticed that parameter $c_i$ is related to Prandtl/Schmidt number in Kolmogorov case but this definite relation is broken in non-Kolmogorov cases because of the presence of inhomogeneous, anisotropic and/or underdeveloped turbulence. In what follows, we consider $c_i$ as a direct parameter, and set its range in Appendix I.
	
\subsection{Non-Kolmogorov temperature-salinity co-spectrum}
In the Kolmogorov case the models of temperature-salinity co-spectrum have been obtained by analogy with the single scalar (temperature/salinity)  spectrum \cite{Nikishov_2000,Yao_17,Yi_18,Yao_19} but such an analogy is unavailable if the exponents of temperature and salinity spectra are different. 
Hence, for the non-Kolmogorov case, the temperature-salinity co-spectrum should be obtained by other means. 
In this section, we will derive the temperature-salinity co-spectrum based on the Upper-Bound Limitation \cite{Bernard1975,Washburn1996,Seim1999} and the concept of spectral correlation \cite{Nash1999}.

As proven in Section 5.2.5 of \cite{Bernard1975}, the Upper-Bound Limitation gives the relation between scalar spectra ${\phi _T}$, ${\phi _S}$ and their co-spectrum ${\phi _{TS}}$:
\begin{equation}
0 \le {\phi _{TS}}\left( \kappa  \right) \le {\left[ {{\phi _T}\left( \kappa  \right){\phi _S}\left( \kappa  \right)} \right]^{1/2}}.
\label{eq21}
\end{equation}
In Ref. \cite{ross_2004} the Upper-Bound Limitation was extended to three-dimensional case:
\begin{equation}
0 \le {\Phi _{TS}}\left( \kappa,\alpha_T,\alpha_S  \right) \le {\left[ {{\Phi _T}\left( \kappa,\alpha_T  \right){\Phi _S}\left( \kappa,\alpha_S  \right)} \right]^{1/2}}.
\label{eq22}
\end{equation}
By adopting the concept of spectral correlation \cite{footnote1,Nash1999}, and combining Eq.(\ref{eq13}) with Eq. (\ref{eq22}), we obtain the temperature-salinity co-spectrum as
\begin{equation}
\begin{split}
{\Phi _{TS}}\left( \kappa,\alpha_T,\alpha_S  \right)& = {\gamma _{ST}}\left( {\kappa \eta } \right){\left[ {{\Phi _T}\left( {\kappa ,{\alpha _T}} \right){\Phi _S}\left( {\kappa ,{\alpha _S}} \right)} \right]^{1/2}}\\
&= {\gamma _{ST}}\left( {\kappa \eta } \right){A_{TS}}\left( {{\alpha _T},{\alpha _S}} \right)C_{TS}^{\rm{2}}{\kappa ^{ - ({\alpha _T} + {\alpha _S})/2}}{g_{TS}}\left( {\kappa \eta } \right) ,
\end{split}
\label{eq23}
\end{equation}
with
\begin{equation}
C_{TS}^{\rm{2}} = {\left( {C_T^{\rm{2}}C_S^{\rm{2}}} \right)^{1/2}},
\label{eq24}
\end{equation}
\begin{equation}
{A_{TS}}\left({\alpha _T},{\alpha _S}\right) = {\left[ {A\left( {{\alpha _T}} \right)A\left( {{\alpha _S}} \right)} \right]^{1/2}},
\label{eq25}
\end{equation}
\begin{equation}
{g_{TS}}\left(\kappa\eta\right) = {\left[ {{g_T}\left( {\frac{{\kappa \eta }}{{h\left( {{\alpha _T},{c_T}} \right)}}} \right){g_S}\left( {\frac{{\kappa \eta }}{{h\left( {{\alpha _S},{c_S}} \right)}}} \right)} \right]^{1/2}},
\label{eq26}
\end{equation}
where ${\gamma _{ST}}\left( {\kappa \eta } \right)$ is a correlation factor describing the degree of correlation between temperature spectrum and salinity spectrum, and $0 \le {\gamma _{ST}}\left( {\kappa \eta } \right) \le 1$. 
When $\gamma_{ST} = 1$, Eq. (23) refers to a fully correlated co-spectrum; when $\gamma_{ST} = 0$, Eq.(23) refers to an uncorrelated co-spectrum that ${\Phi _{TS}}=0$; when $0<\gamma_{ST}<1$, the co-spectrum is partially correlated. Details about partially correlated co-spectrum are given as follows.

\begin{figure}
	\centering
	\subfigure[$\alpha_{T} = 11/3$, $c_T = 2.2\times 10^{-3}$;]{
		\centering
		\includegraphics[width=0.28\linewidth]{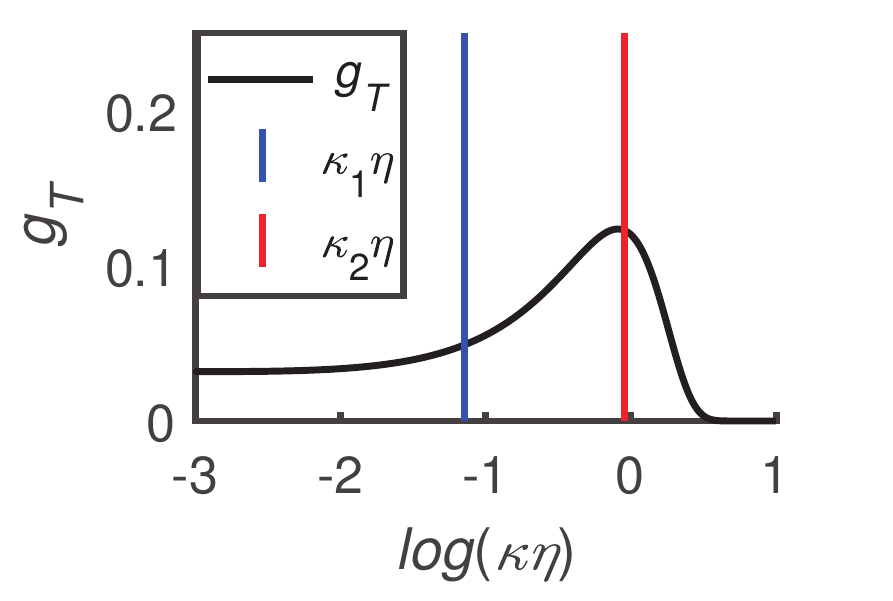}	
	}
	\subfigure[$\alpha_{T} = 12.5/3$, $c_T = 2.2\times 10^{-3}$;]{
		\centering
		\includegraphics[width=0.28\linewidth]{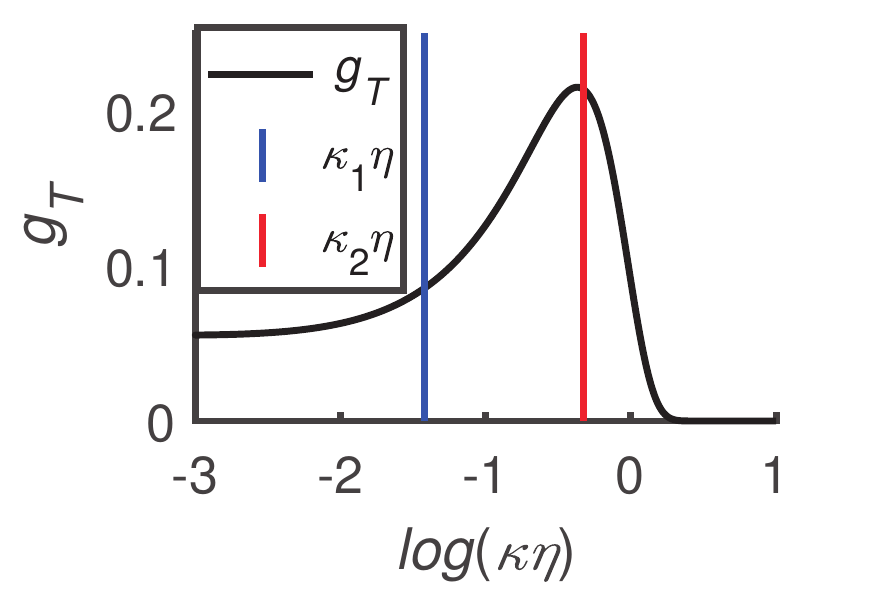}	
	}
	\subfigure[$\alpha_{T} = 14/3$, $c_T = 2.2\times 10^{-3}$;]{
		\centering
		\includegraphics[width=0.28\linewidth]{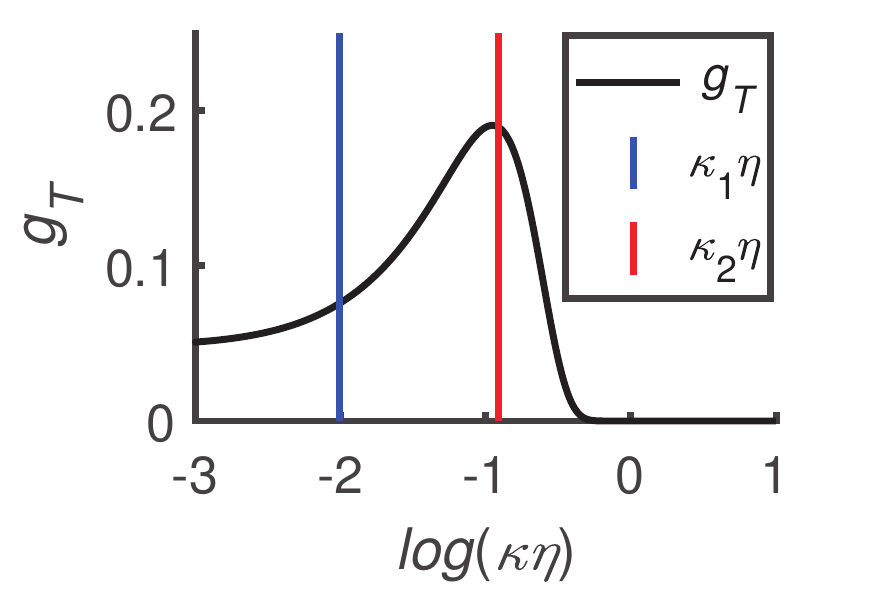}	
	}
	\\
	\subfigure[$\alpha_{T} = 14/3$, $c_T = 3.6\times 10^{-3}$;]{
		\centering
		\includegraphics[width=0.28\linewidth]{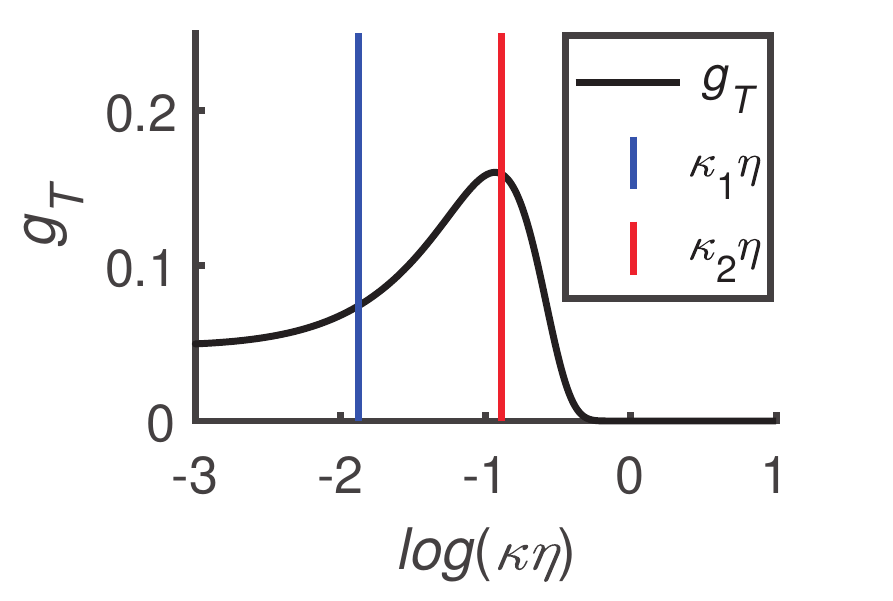}	
	}
	\subfigure[$\alpha_{T} = 14/3$, $c_T = 2.2\times 10^{-3}$;]{
		\centering
		\includegraphics[width=0.28\linewidth]{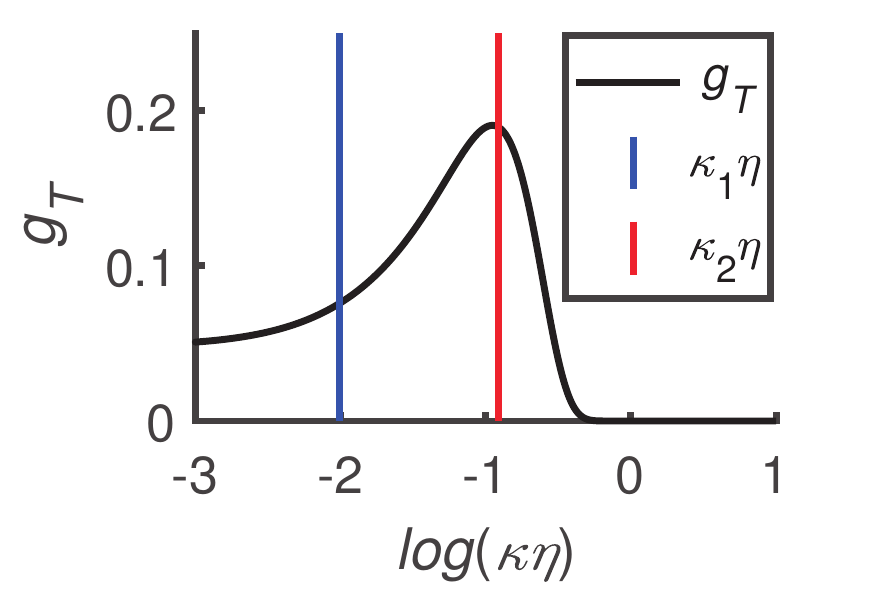}	
	}
	\subfigure[$\alpha_{T} = 14/3$, $c_T = 1.5\times 10^{-3}$;]{
		\centering
		\includegraphics[width=0.28\linewidth]{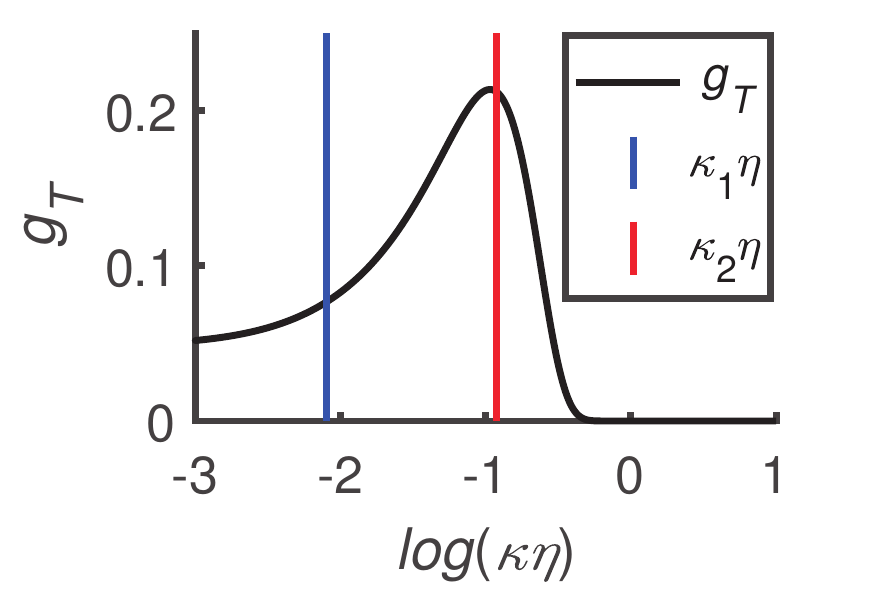}	
	}
	\caption{The locations of $\kappa_{1}\eta$ (`\textcolor{blue}{|}') and $\kappa_{2}\eta$ (`\textcolor{red}{|}') defined by Eqs. (\ref{eq29})-(\ref{eq30}). Here the horizontal and vertical axes are all dimensionless.}
	\label{fig:2}
\end{figure}

According to the concept of spectral correlation \cite{Nash1999,ross_2004}, temperature fluctuation $T'$ and salinity fluctuation $S'$ are highly correlated if they are both driven by eddy diffusion, but the correlation will be broken down if $T'$ is driven by temperature molecular diffusion. Hence, the following should hold \cite{footnote5}:
\begin{itemize}
	\item When ${\kappa}$ belongs to \textbf{the inertial-convective range of $\Phi _{T}$} (i.e. $g_T \propto \kappa^{0}$),
	the salinity spectrum is generally in its inertial-convective range \cite{footnote3}.
	Thus, both $T'$ and $S'$ are governed by eddy diffusion, and they have \textbf{a high correlation}, i.e. $\gamma_{ST} = \gamma_{\max } \le 1$.
	
	\item When ${\kappa}$ belongs to \textbf{the viscous-convective range of $\Phi _{T}$ } (i.e. $g_T \propto \kappa^{2/3}$),
	$T'$ is consumed by viscosity but $S'$ is still governed by eddy diffusion.
	The \textbf{correlation} begins to decrease in this range, and it has been observed in \cite{Yeung_2000} that correlation decreases monotonically. Hence, we have $\mathrm{d}{\gamma _{ST}}/\mathrm{d} \kappa \le0$.
	
	\item When ${\kappa}$ belongs to \textbf{the viscous-diffusive range of $\Phi _{T}$} (i.e. $g_T$ decreases fast with $\kappa$),
	$T'$ is primarily depleted by temperature molecular diffusion, which leads to a \textbf{very low correlation} between $T'$ and $S'$, i.e. ${\gamma _{ST}} \approx {\rm{0}}$.
\end{itemize}
Thus the value of correlation parameter ${\gamma _{ST}}\left( \kappa \eta \right)$ obeys the following constraints:
\begin{equation}
\left\{{\begin{array}{*{20}{l}}
{{\gamma _{ST}}\left( {\kappa \eta } \right) = {\gamma _{\max }}\le 1}&{{\rm{when }}\ \kappa  \ll \kappa _1,}\\
{{\gamma _{ST}}\left( {\kappa \eta } \right) \in \left[ {0 ,{\gamma _{\max }}} \right]\ {\rm{and}}\ {\mathrm{d}{\gamma _{ST}}/\mathrm{d} \kappa} \le 0}&{{\rm{when }}\ \kappa _1 \ll \kappa  \ll {\kappa _2},}\\
{{\gamma _{ST}}\left( {\kappa \eta } \right) \approx 0}&{{\rm{when }}\ \kappa  \gg {\kappa _2},}
\end{array}}\right.
\label{eq27}
\end{equation}
where $\kappa_1$ defines the transition between inertial-convective and viscous-convective ranges of $\Phi_T$, 
$\kappa_2$ defines the transition between viscous-convective and viscous-diffusive ranges of $\Phi_T$.
According to \cite{Hill1978}, we have the following defining relations for $\kappa_1$ and $\kappa_2$ in H4-based non-Kolmogorov model:
\begin{equation}
\frac{{{\kappa _1}\eta }}{{h\left( {{\alpha _T}},{c_T} \right)}} = a,
\label{eq28}
\end{equation}
and
\begin{equation}
\frac{{{\kappa _2}\eta }}{{h\left( {{\alpha _T}},{c_T} \right)}} = {\left( {\frac{{3{a^{4/3}}}}{{22Q{c_T}}}} \right)^{1/2}},
\label{eq29}
\end{equation}
where $\eta$ is the Kolmogorov scale; $h\left(\alpha_T,c_T\right)$ is the non-Kolmogorov scaling function given in Eq. (\ref{eq18}); 
$a$ is a constant approximating to 0.072 \cite{Nikishov_2000}; $Q$ is another constant about 2.35 \cite{Nikishov_2000}; and $c_T$ has been given in Eq. (\ref{eq6}).
The locations of $\kappa_1\eta$ and $\kappa_2\eta$ are marked by `\textcolor{blue}{|}' and `\textcolor{red}{|}' in Fig.\ref{fig:2}, respectively, and `---' refers to $g_T$.
It shows that $\kappa_1\eta$ and $\kappa_2\eta$ mark the transitions between different ranges very well.

For mathematical simplicity of discussion, we suppose that the correlation factor in \textit{fully correlated} case is
\begin{equation}
\gamma_{ST}=1,
\label{eq30}
\end{equation}
and in \textit{partially correlated} case it takes form 
\begin{equation}
{\gamma _{{\rm{S}}T}}\left( {\kappa \eta } \right) = \frac{{1 - \tanh \left\{ {\left[ {\log (\kappa \eta ) - \left( {\log ({\kappa _1}\eta ) + \log ({\kappa _2}\eta )} \right)/2} \right]\rho } \right\}}}{2}{\gamma _{\max }},
\label{eq31}
\end{equation}
with
\begin{equation}
\rho  = \frac{{2p}}{{\log ({\kappa _2}\eta ) - \log ({\kappa _1}\eta )}},\quad \gamma_{\max }=1,
\label{eq32}
\end{equation}
where $p$ controls the transition speed of $\gamma_{ST}$ from $\gamma_{\max }$ to $0$.

\begin{figure*}[!t]
	\centering
	\begin{minipage}[c]{.43\textwidth}
		\subfigure[$q(\kappa\eta)$;]{
			\centering
			\includegraphics[width=1\textwidth]{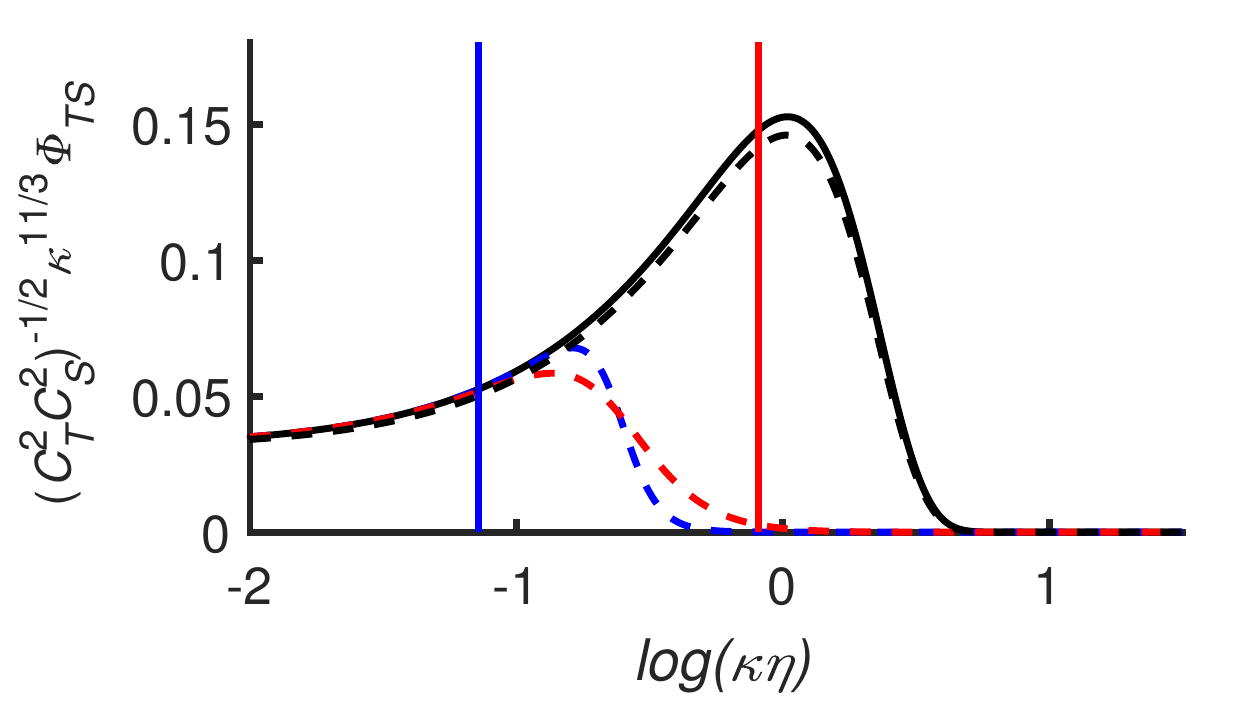}	
		}
	\end{minipage}%
	\begin{minipage}[c]{.15\textwidth}
	\centering
	\includegraphics[width=1\textwidth]{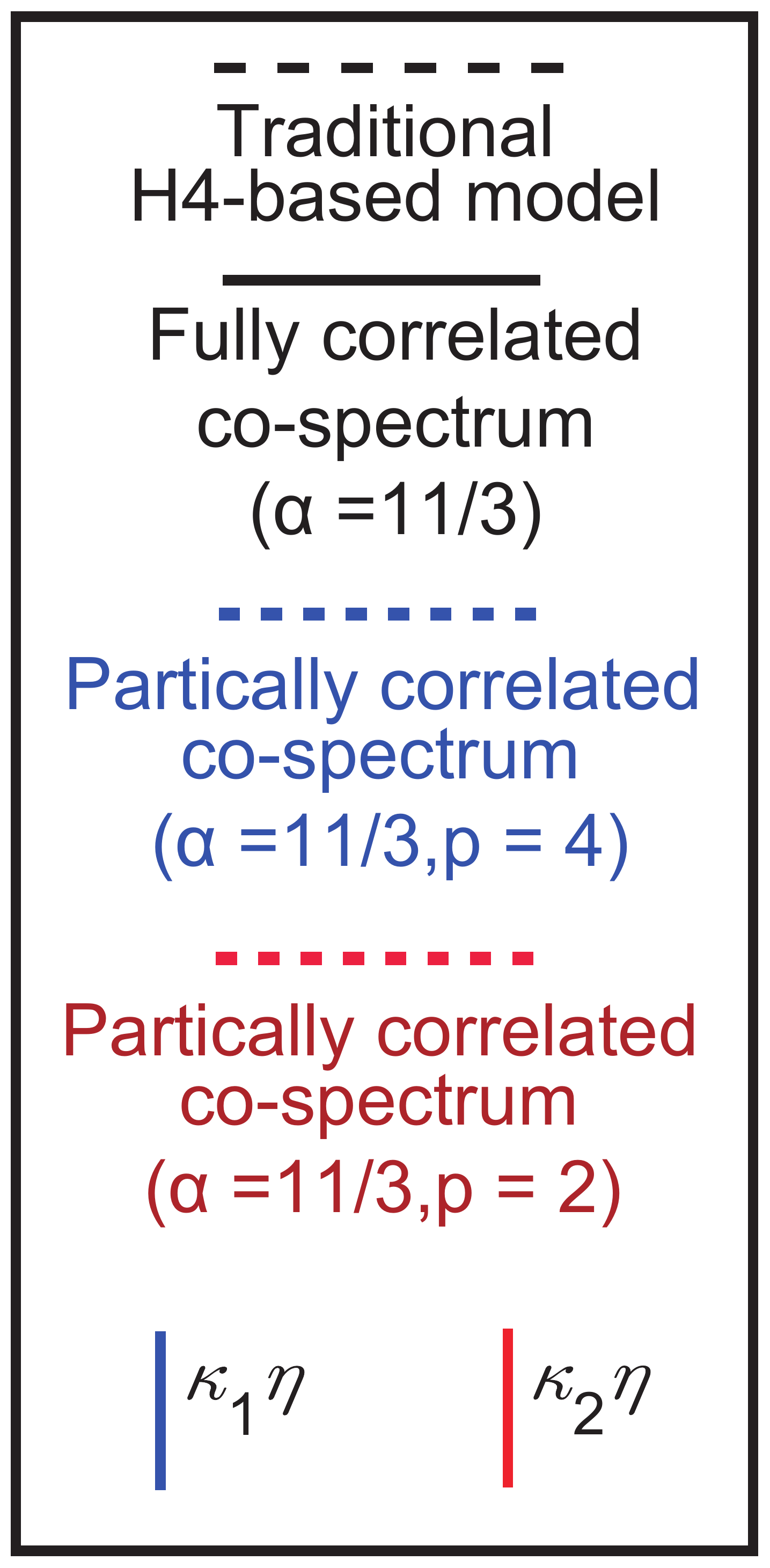}\\
	$\,$
	\end{minipage}
	\begin{minipage}[c]{.39\textwidth}
		\subfigure[$\gamma_{ST}(\kappa\eta)$;]{
			\centering
			\includegraphics[width=1\textwidth]{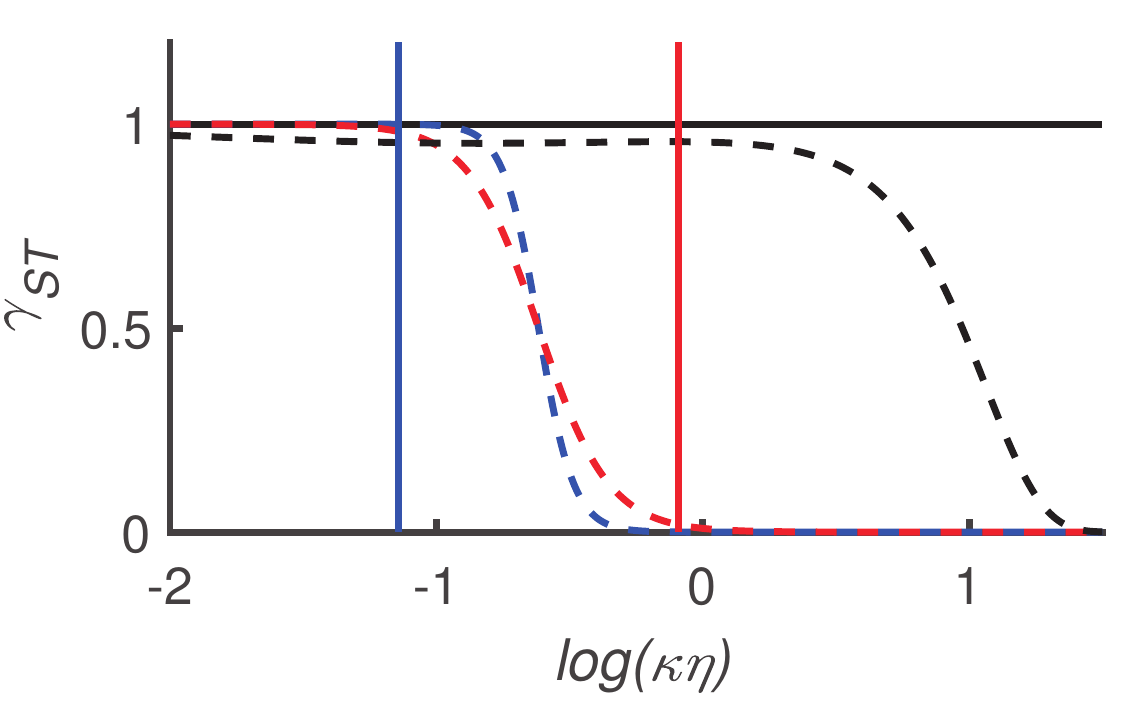}	
		}
	\end{minipage}%
	\caption{Comparing (a) the non-dimensional function $q(\kappa\eta)$ and (b) the correlation factor $\gamma_{ST}(\kappa\eta)$ corresponding to the proposed co-spectrum with these corresponding to conventional co-spectrum \cite{Yao_19}. Values of parameters are listed in Appendix II.  The unit of the vertical axis of (a) is $[\rm m^{(\alpha_T+\alpha_S)/2-11/3}]$. $\gamma_{ST}$ and $\kappa\eta$ are dimensionless.}
	\label{fig:3}
\end{figure*}

In Figure \ref{fig:3} we compare Eq. (\ref{eq23}) with the conventional co-spectrum \cite{Yao_19} limiting ourselves  to \textit{Kolmogorov case} ($\alpha_S = \alpha_T = 11/3$). Fig. \ref{fig:3}(a) shows non-dimensional function $q(\kappa\eta) = {{( {C_T^{\rm{2}}C_S^{\rm{2}}} )}^{ - 1/2}}{\kappa ^{11/3}}{\Phi _{TS}}$ varying with $\log\left(\kappa\eta\right)$,
where `\textcolor{black}{-{}-{}-}' refers to the traditional co-spectrum \cite{Yao_19};
`\textcolor{black}{---}' refers to the proposed co-spectrum in Eq. (\ref{eq23}) with a full correlation $\gamma_{ST} = 1$; 
the curves `\textcolor{blue}{-{}-{}-}' and `\textcolor{red}{-{}-{}-}' refer to the partially correlated co-spectra with $p = 4$ and $p = 2$, respectively. 
The vertical lines `\textcolor{blue}{|}' and `\textcolor{red}{|}' mark the locations of $\kappa_1\eta$ and $\kappa_2\eta$, respectively. 
With similar legends, Fig. \ref{fig:3}(b) shows correlation factor $\gamma_{ST}$ varying with $\log\left(\kappa\eta\right)$ \cite{footnote2}.

\begin{figure*}
	\centering
	\begin{minipage}[c]{.84\textwidth}
	\subfigure[$\alpha_{T} = 14/3$, $\alpha_{S} = 11/3$;]{
		\centering
		\includegraphics[width=0.3\textwidth]{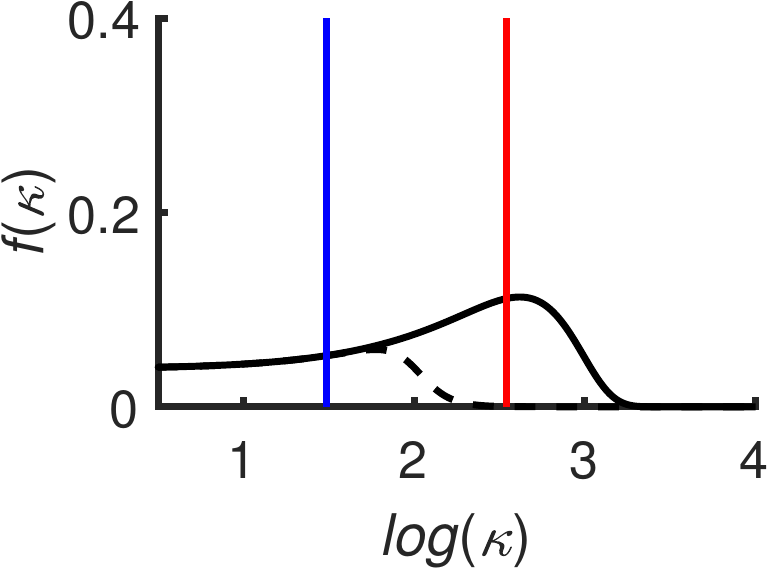}	
	}
	\subfigure[$\alpha_{T} = 14/3$, $\alpha_{S} = 12.5/3$;]{
		\centering
		\includegraphics[width=0.3\textwidth]{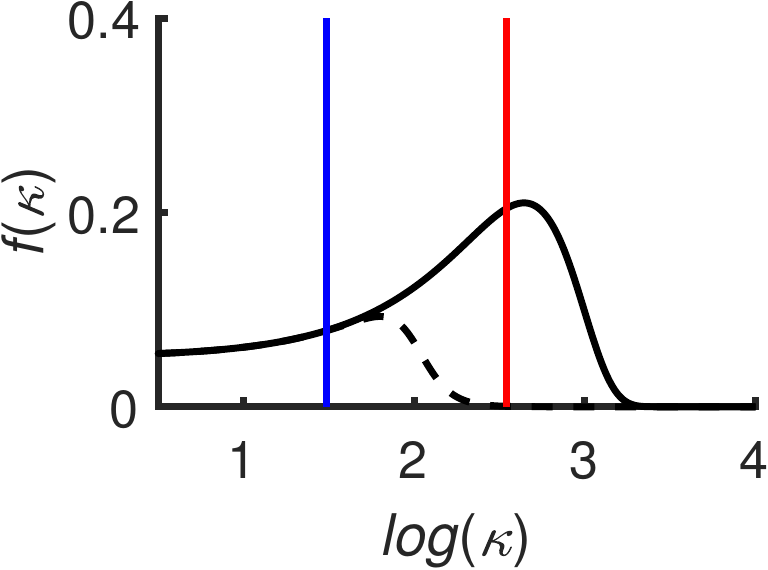}	
	}
	\subfigure[$\alpha_{T} = 14/3$, $\alpha_{S} = 14/3$;]{
		\centering
		\includegraphics[width=0.3\textwidth]{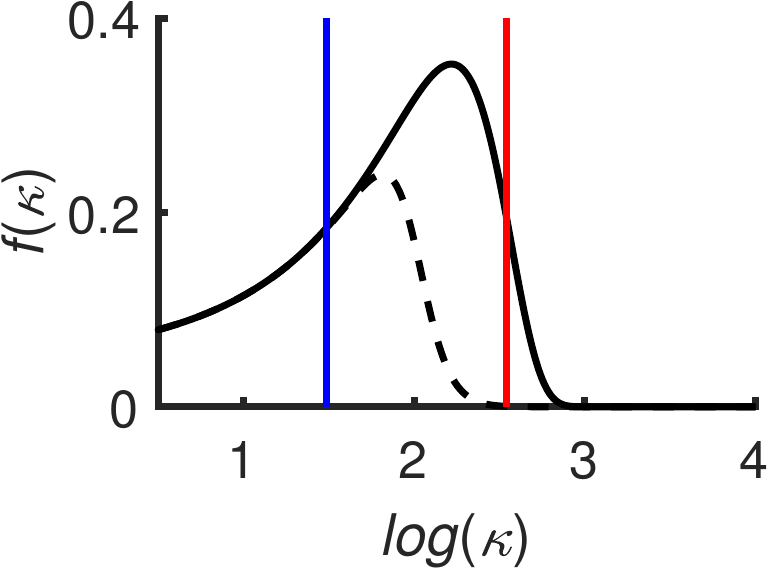}	
	}
	\\
	\subfigure[$\alpha_{T} = 11/3$, $\alpha_{S} = 12/3$;]{
		\centering
		\includegraphics[width=0.3\textwidth]{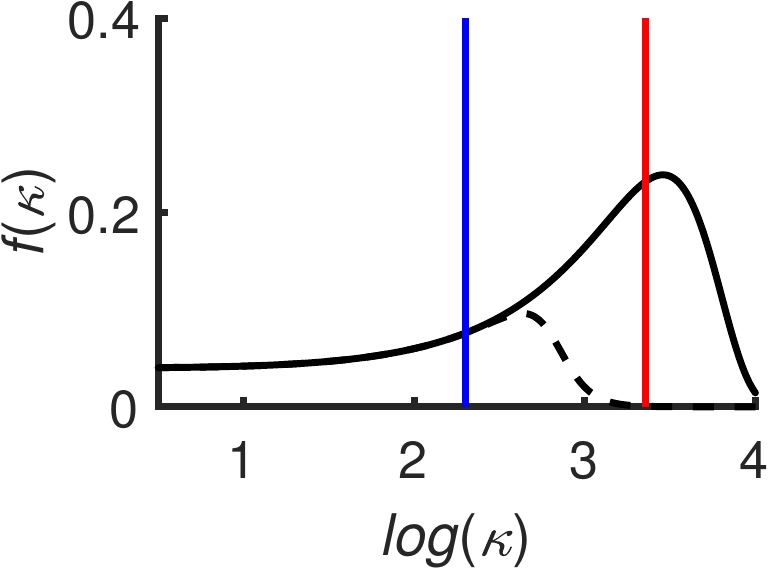}	
	}
	\subfigure[$\alpha_{T} = 12.5/3$, $\alpha_{S} = 12/3$;]{
		\centering
		\includegraphics[width=0.3\textwidth]{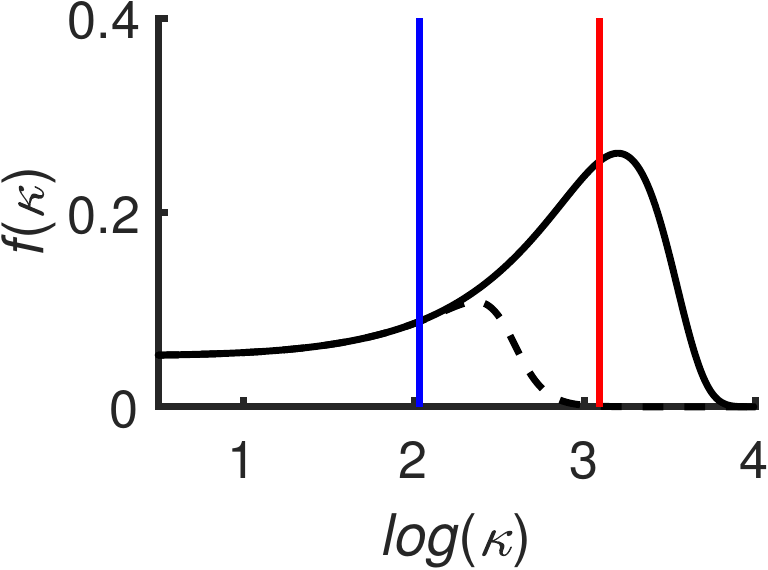}	
	}
	\subfigure[$\alpha_{T} = 14/3$, $\alpha_{S} = 12/3$;]{
		\centering
		\includegraphics[width=0.3\textwidth]{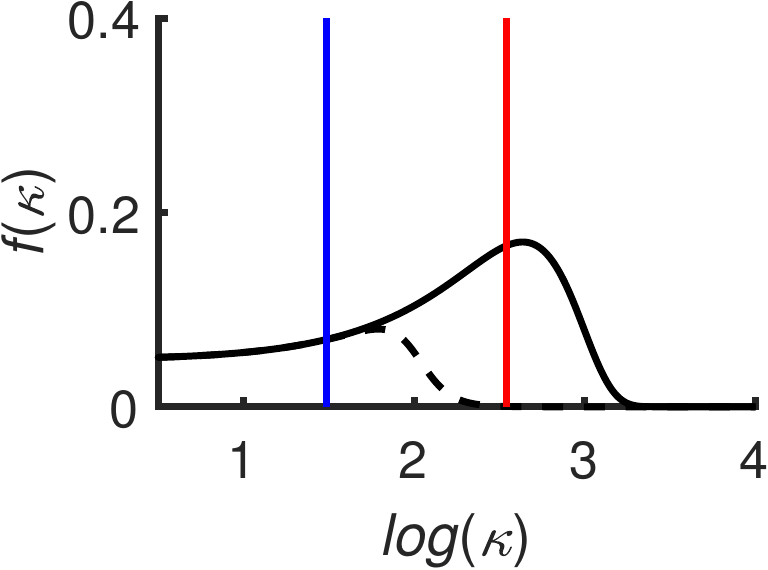}	
	}
	\end{minipage}%
	\begin{minipage}[c]{.13\textwidth}
		\centering
		\includegraphics[width=1\textwidth]{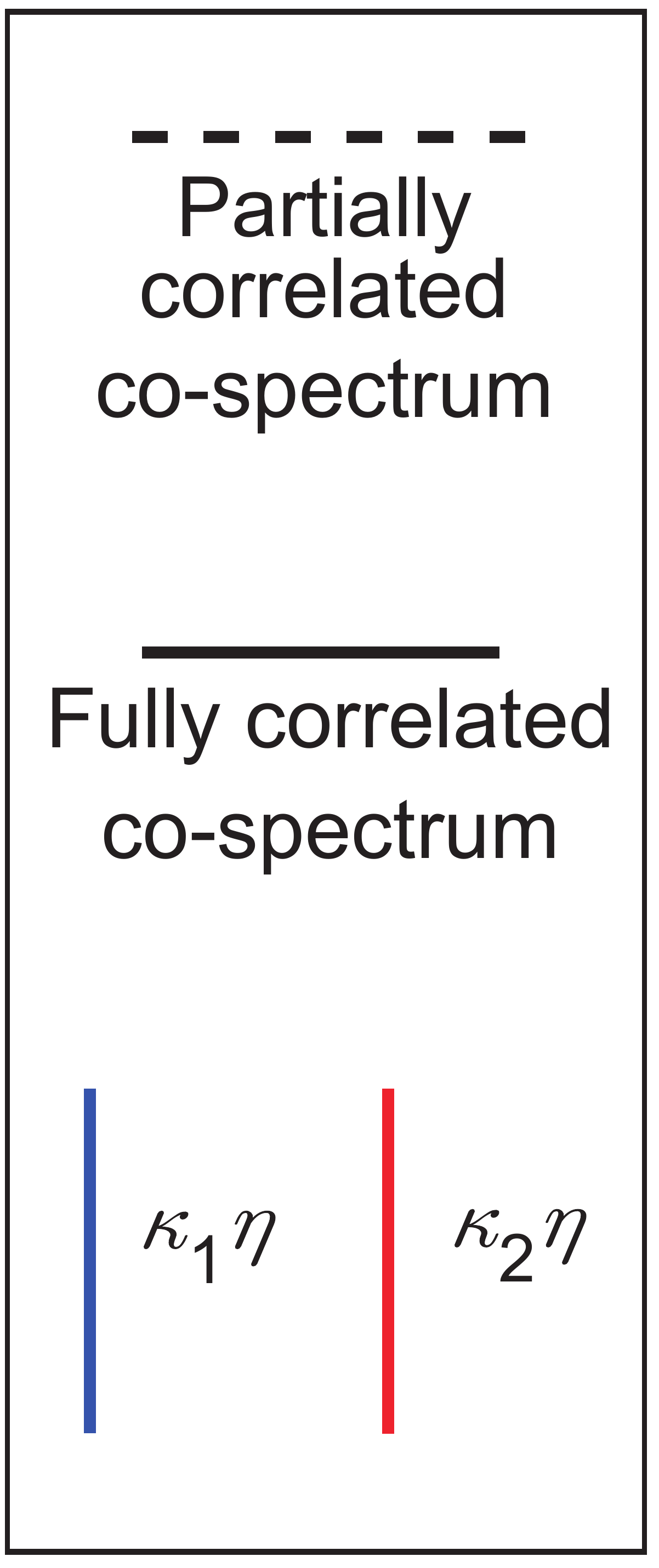}	\\
		$\,$\\
		$\,$
	\end{minipage}
	\caption{The curves of the dimensionless function $f(\kappa) = {\kappa ^{(\alpha_T+\alpha_S)/2}}{{( {C_T^{\rm{2}}C_S^{\rm{2}}} )}^{ - 1/2}}{\Phi _{TS}}$ with different values of $\alpha_{T}$ and $\alpha_{S}$. Here $\kappa$ is dimensionless.}
	\label{fig:4}
\end{figure*}

Figure \ref{fig:3} shows that: \textit{for the Kolmogorov case}
and in comparison with the conventional co-spectrum [\citen{Yao_19}], 
the proposed partially correlated co-spectrum has a higher correlation in the temperature inertial-convective range ($\kappa \ll {\kappa _{1}}$), 
a lower correlation in the temperature viscous-convective range ($\kappa_1 \ll \kappa \ll {\kappa _{2}}$),  
and also a low correlation in the temperature viscous-diffusive range ($\kappa  \gg {\kappa _{2}}$). 
Furthermore, Fig.\ref{fig:3} (a) indicates that the proposed fully correlated co-spectrum tends to the conventional co-spectrum when $\alpha_{T} = \alpha_{S} = 11/3$.

To examine the co-spectrum \textit{in non-Kolmogorov case}, 
and to verify its de-correlation within temperature viscous-convective range,
we plot $log$ of non-dimensional function $f(\kappa) = {\kappa ^{(\alpha_T+\alpha_S)/2}}{{( {C_T^{\rm{2}}C_S^{\rm{2}}} )}^{ - 1/2}}{\Phi _{TS}}$ in Fig.\ref{fig:4}, and compare the fully correlated co-spectrum (`---') with the partially correlated co-spectrum (`-{}-{}-') at $p = 3$.
Same as before, $\kappa_1\eta$ and $\kappa_2\eta$ are marked by `\textcolor{blue}{|}' and `\textcolor{red}{|}', respectively. It is shown that the proposed co-spectrum has low correlation in the temperature inertial-convective range, as expected. This agrees with Eq. (\ref{eq27}).

\textit{Thus we have obtained a temperature-salinity co-spectrum with a non-Kolmogorov power law $(\alpha_{T}+\alpha_{S})/2$ and a flexible correlation factor $\gamma_{ST}$ [see Eq. (\ref{eq23})].} If $\gamma_{ST} = 1$, the proposed co-spectrum is fully correlated, and it approximately reduces to the conventional co-spectrum when $\alpha_{T}=\alpha_{S}=11/3$; if $\gamma_{ST} = 0$, the proposed co-spectrum is uncorrelated, i.e. ${\Phi _{TS}}=0$; if $\gamma_{ST}$ obeys Eq. (\ref{eq31}), the proposed co-spectrum is partially correlated.
As we expected, the new co-spectrum model has a power law between $\alpha_{T}$ and $\alpha_{S}$; 
if the temperature and salinity fields are both Kolmogorov ($\alpha_{T} = \alpha_{S} = 11/3$), the co-spectrum is also Kolmogorov ($\alpha_{TS} = 11/3$).

\section{OTOPS with anisotropy and non-Kolmogorov power law}
In general, the oceanic refractive-index fluctuation $n'$ is approximately given by a linear combination of temperature fluctuation $T'$ and salinity fluctuation $S'$ \cite{Nikishov_2000,KOROTKOVA2020_1,footnote4}:
\begin{equation}
n' \approx n'_T T' + n'_S S',
\label{eq33}
\end{equation}
with
\begin{equation}
n'_T = \frac{{dn'}}{{dT'}}, \quad
n'_S = \frac{{dn'}}{{dS'}}. 
\label{eq34}
\end{equation}
This implies that the spectrum of $n'$ is approximately given by linear combination
\begin{equation}
{\Phi _{n0}}( \kappa) = n{'_T}^2 \Phi_T(\kappa) + n{'_S}^2 \Phi_S(\kappa) + 2{n'_T}{n'_S} \Phi_{TS}(\kappa),
\label{eq35}
\end{equation}
where $\Phi_T$ is the temperature spectrum, $\Phi_S$ is the salinity spectrum, and $\Phi_{TS}$ is the temperature-salinity co-spectrum. On incorporating Eqs. (\ref{eq13}) and (\ref{eq23}) into Eq. (\ref{eq35}), we obtain the following expression for the NK-OTOPS: 
\begin{equation}
\begin{split}
{\Phi _{n0}}\left( \kappa  \right)&= n{'_T}^2{\Phi _T}\left( \kappa  \right) + n{'_S}^2{\Phi _S}\left( \kappa  \right) + 2n{'_T}n{'_S}{\gamma _{ST}}\left( {\kappa \eta } \right)\sqrt {{\Phi _T}\left( \kappa  \right){\Phi _S}\left( \kappa  \right)}\\
&= n{'_T}^2C_T^{\rm{2}}A\left( {{\alpha _T}} \right){\kappa ^{ - {\alpha _T}}}{g_T}(\kappa \eta/h_T) + {n{'_S}}^2C_S^{\rm{2}}A\left( {{\alpha _S}} \right){\kappa ^{ - {\alpha _S}}}{g_S}(\kappa \eta/h_S) + 2{n'_T}{n'_S}{\gamma _{ST}}\\
&\quad{\left( {C_T^{\rm{2}}C_S^{\rm{2}}} \right)^{1/2}}{\left[ {A\left( {{\alpha _T}} \right)A\left( {{\alpha _S}} \right)} \right]^{1/2}}{\kappa ^{ - ({\alpha _T} + {\alpha _S})/2}}{\left[ {{g_T}(\kappa \eta/h_T){g_S}(\kappa \eta/h_S)} \right]^{1/2}}.
\end{split}
\label{eq36}
\end{equation}
with
\begin{equation}
h_T = h\left(\alpha_T,c_T\right),\quad h_S = h\left(\alpha_S,c_S\right).
\end{equation}

To make the developed NK-OTOPS more physical we now implement the finite outer-scale cut-off and extend it to the anisotropic case. To obtain the first extension we use the filter function with exponential form \cite{Voitsekhovich,Li_19}:
\begin{equation}
{\Phi _{n{\rm{1}}}}\left( \kappa  \right) = \left[ {1 - \exp \left( { - \frac{{{\kappa ^2}}}{{{\kappa _0}^2}}} \right)} \right]{\Phi _{n{\rm{0}}}}\left( \kappa  \right),
\label{eq37}
\end{equation}
where $\kappa_0$ is the outer-scale cut-off wavenumber defined by $\kappa_0 \approx 4\pi/L_0$ with $L_0 (\rm{m})$ representing the outer scale. Further the \textit{anisotropic} NK-OTOPS can be obtained on following \cite{Toselli_2014} as:
\begin{equation}
{\Phi _{n2}}\left( \bm{\kappa}  \right) = \mu^2{\Phi _{n1}}\left( {{\kappa _{{\rm{iso}}}}} \right),
\label{eq38}
\end{equation}
where $ \mu $ is the anisotropy parameter, $\bm{\kappa }$ is the three-dimensional wavenumber, and $\bm{\kappa }_{\rm{iso}}$ is a isotropisizing transformation of $\bm{\kappa }$:
\begin{equation}
\bm{\kappa } = {\left( { {\kappa _x},{\kappa _y},{\kappa _z}} \right)^{\rm{T}}},
\quad
\bm{\kappa }_{\rm{iso}} = {\left( {\mu {\kappa _x},\mu {\kappa _y},{\kappa _z}} \right)^{\rm{T}}},
\quad
{\kappa _{{\rm{iso}}}} = \left| \bm{\kappa }_{\rm{iso}}\right|,
\label{eq39}
\end{equation}
$T$ is denoting vector transpose.

Thus in this section, \textit{a non-Kolmogorov OTOPS (NK-OTOPS) is given in Eq. (\ref{eq36}), while its extended form for outer-scaled and anisotropic cases are presented by Eqs. (\ref{eq37}) and (\ref{eq38}), respectively}.

\section{Spherical wave propagation in oceanic optical turbulence}
As an example of applying the NK-OTOPS, and on taking into account the significance of the spherical wave statistics for the extended Huygens-Fresnel principle, we will calculate and analyze the 2nd-order statistics of a spherical wave. In particular, 
in Section 4.1, the wave structure function (WSF) of a spherical wave in oceanic turbulence will be derived; in Section 4.2, the vector and scalar versions of the coherence radius will be defined and examined; and in Section 4.3, the co-effect of temperature and salinity on spherical wave's propagation will be discussed by calculating its scalar coherence radius varying with $\alpha_T$, $\alpha_S$, $c_T$ and $c_S$.

\subsection{2nd-order statistical moments and wave structure function of spherical wave}
\subsubsection{A. 2nd-order statistical moments}
\textit{We will first derive the 2nd-order statistical moment of a spherical wave propagating in the non-Kolmogorov oceanic optical turbulence}. According to Eq. (59) of chapter 5 in \cite{Phillips2005Laser}, for horizontal channels (along y-axis) this quantity has form:
\begin{equation}
{E_{2\_h}}\left( {{\bm{r}_1},{\bm{r}_2}} \right) = \frac{{2\pi {k^2}}}{{n_0^2}}\int_0^L {d\eta \int {\int_{ - \infty }^{ + \infty } {{d^2}\bm{\kappa}  \cdot {\Phi _{n2}}(\bm{\kappa} )\exp \left[ {i\bm{\kappa} \left( {\gamma {\bm{r}_1} - {\gamma ^*}{\bm{r}_2}} \right) - \frac{{i{\bm{\kappa} ^2}}}{{2k}}\left( {\gamma  - {\gamma ^*}} \right)\left( {L - \eta } \right)} \right]} } },
\label{eq40}
\end{equation}
with
\begin{equation}
\bm{\kappa}  = {\left( {{\kappa _x},{\kappa _z}} \right)^{\rm{T}}},
{\bm{r}_1} = {\left( {{r_{1x}},{r_{1z}}} \right)^{\rm{T}}},
{\bm{r}_2} = {\left( {{r_{2x}},{r_{2z}}} \right)^{\rm{T}}},
\label{eq41}
\end{equation}
where $L$ is the propagation distance from the source plane, $k$ is the wavenumber defined as $2\pi n_0/\lambda$, $n_0$ being the average refractive-index, $\Phi_{n2}$ is the anisotropic NK-OTOPS as given by Eq. (\ref{eq38}). For a spherical wave, $\gamma = \gamma^* = 1$. On assuming that
\begin{equation}
\quad {\bm{\kappa} _{{\rm{t}}}} = {\left( {\mu {\kappa _x},{\kappa _z}} \right)^{\rm{T}}},
\quad{\bm{r}_{1\_{\rm{iso}}}} = {\left( {{r_{1x}}/\mu ,{r_{1z}}} \right)^{\rm{T}}},
\quad{\bm{r}_{2\_{\rm{iso}}}} = {\left( {{r_{2x}}/\mu ,{r_{2z}}} \right)^{\rm{T}}},
\label{eq42}
\end{equation}
and combining Eq. (\ref{eq38}) with Eq. (\ref{eq40}), we get
\begin{equation}
\begin{split}
{E_{2\_h}}\left( {{\bm{r}_1},{\bm{r}_2}} \right) &= \frac{2\pi {k^2}}{n_0^2\mu}\int_0^L {d\eta \int {\int_{ - \infty }^{ + \infty } {{d^2}{\bm{\kappa} _{{\rm{t}}}} \cdot {\mu ^2}{\Phi _{n1}}({\bm{\kappa} _{{\rm{t}}}})\exp \left[ {i{\bm{\kappa} _{{\rm{t}}}}\left( {{\bm{r}_{1\_{\rm{iso}}}} - {\bm{r}_{2\_{\rm{iso}}}}} \right)} \right]} } }\\
&= \frac{4{\pi ^2}{k^2}\mu L}{n_0^2}\int_0^{ + \infty } {d{\kappa _{{\rm{t}}}} \cdot {\kappa _{{\rm{t}}}}{\Phi _{n1}}({\kappa _{{\rm{t}}}}){J_0}\left[ {{\kappa _{{\rm{t}}}}\left| {{\bm{r}_{1\_{\rm{iso}}}} - {\bm{r}_{2\_{\rm{iso}}}}} \right|} \right]},
\end{split}
\label{eq43}
\end{equation}
where $\Phi_{n1}$ is the outer-scaled NK-OTOPS in Eq. (\ref{eq37}).
On setting $\bm{ \rho } = \bm{r}_1 - \bm{r}_2$, we find that the 2nd-order statistical moment of a spherical wave along a horizontal channel (along the y-axis) becomes
\begin{equation}
{E_{2\_h}}\left( \bm{\rho}  \right) = \frac{4{\pi ^2}{k^2}\mu L}{n_0^2}\int_0^{ + \infty } {d{\kappa _{{\rm{t}}}} \cdot {\kappa _{{\rm{t}}}}{\Phi _{n1}}({\kappa _{{\rm{t}}}}){J_0}\left[ {{\kappa _{{\rm{t}}}}\sqrt {{\mu ^{-2}}\rho _x^2 + \rho _z^2} } \right]},
\label{eq44}
\end{equation}
where
\begin{equation}
{\kappa _{\rm{t}}} = \left| {{{\left( {\mu {\kappa _x},{\kappa _z}} \right)}^{\rm{T}}}} \right|.
\label{eq45}
\end{equation}
Similarly, the 2nd-order statistical moment of a spherical wave in a vertical channel (along the z-axis) becomes 
\begin{equation}
{E_{2\_v}}\left( \bm{\rho}  \right) = \frac{4{\pi ^2}{k^2}L}{n_0^2}\int_0^{ + \infty } {d{\kappa _{{\rm{t}}}} \cdot {\kappa _{{\rm{t}}}}{\Phi _{n1}}({\kappa _{{\rm{t}}}}){J_0}\left[ {{\kappa _{{\rm{t}}}}\sqrt {{\mu ^{-2}}\rho _x^2 + {\mu ^{-2}}\rho _y^2} } \right]},
\label{eq46}
\end{equation}
with
\begin{equation}
{\kappa _{\rm{t}}} = \left| {{{\left( {\mu {\kappa _x},\mu {\kappa _y}} \right)}^{\rm{T}}}} \right|.
\label{eq47}
\end{equation}

\subsubsection{B. Wave structure function of spherical wave}
Next, based on the 2nd-order statistical moments given above, we derive the WSF of a spherical wave in the non-Kolmogorov oceanic optical turbulence. According to the expressions in chapter 6 of \cite{Phillips2005Laser} the WSF of a spherical wave has form:
\begin{equation}
\begin{split}
{D_{sp}}\left( {\bm{\rho} ,L} \right) &= {\mathop{\rm Re}\nolimits} \left[ {\Delta \left( {\bm{\rho} ,L} \right)} \right] = {E_2}\left( {{\bm{r}_1},{\bm{r}_1}} \right) + {E_2}\left( {{\bm{r}_2},{\bm{r}_2}} \right) - 2{E_2}\left( {{\bm{r}_1},{\bm{r}_2}} \right)\\
&= 2{E_2}\left( \bm{0} \right) - 2{E_2}\left( \bm{\rho}  \right),
\end{split}
\label{eq48}
\end{equation}
where $E_2$ is the 2nd-order statistical moment of a spherical wave.
In combining with Eq. (\ref{eq44}), we find that the WSF of a spherical wave in a horizontal channel ($\bm{\rho} = (\rho_x,\rho_z)$) takes form
\begin{equation}
{D_{sp\_h}}\left( {\bm{\rho} ,L} \right) = \frac{8{\pi ^2}{k^2}\mu L}{n_0^2}\int_0^{ + \infty } {d{\kappa _{{\rm{iso}}}} \cdot {\kappa _{{\rm{iso}}}}{\Phi _{n1}}({\kappa _{{\rm{iso}}}})\left[ {1 - {J_0}\left( {{\kappa _{{\rm{iso}}}}\sqrt {{\mu ^{-2}}\rho _x^2 + \rho _z^2} } \right)} \right]} .
\label{eq49}
\end{equation}
Similarly, the WSF of a spherical wave in a vertical channel ($\bm{\rho} = (\rho_x,\rho_y)$) becomes
\begin{equation}
{D_{sp\_v}}\left( {\bm{\rho} ,L} \right) = \frac{8{\pi ^2}{k^2}L}{n_0^2}\int_0^{ + \infty } {d{\kappa _{{\rm{iso}}}} \cdot {\kappa _{{\rm{iso}}}}{\Phi _{n1}}({\kappa _{{\rm{iso}}}})\left[ {1 - {J_0}\left( {{\kappa _{{\rm{iso}}}}\sqrt {{\mu ^{-2}}\rho _x^2 + {\mu ^{-2}}\rho _y^2} } \right)} \right]}.
\label{eq50}
\end{equation}
When $\mu = 1$, the WSFs in horizontal and vertical channels are equal and, hence,
\begin{equation}
{D_{sp\_h}}\left( {\bm{\rho} ,L} \right) = {D_{sp\_v}}\left( {\bm{\rho} ,L} \right) = \frac{8{\pi ^2}{k^2}L}{n_0^2}\int_0^{ + \infty } {d\kappa  \cdot \kappa {\Phi _{n1}}(\kappa )\left[ {1 - {J_0}\left( {\kappa \left| \bm{\rho}  \right| } \right)} \right]}.
\label{eq51}
\end{equation}

\begin{figure*}[!h]
	\centering
	\includegraphics[width=0.24\linewidth]{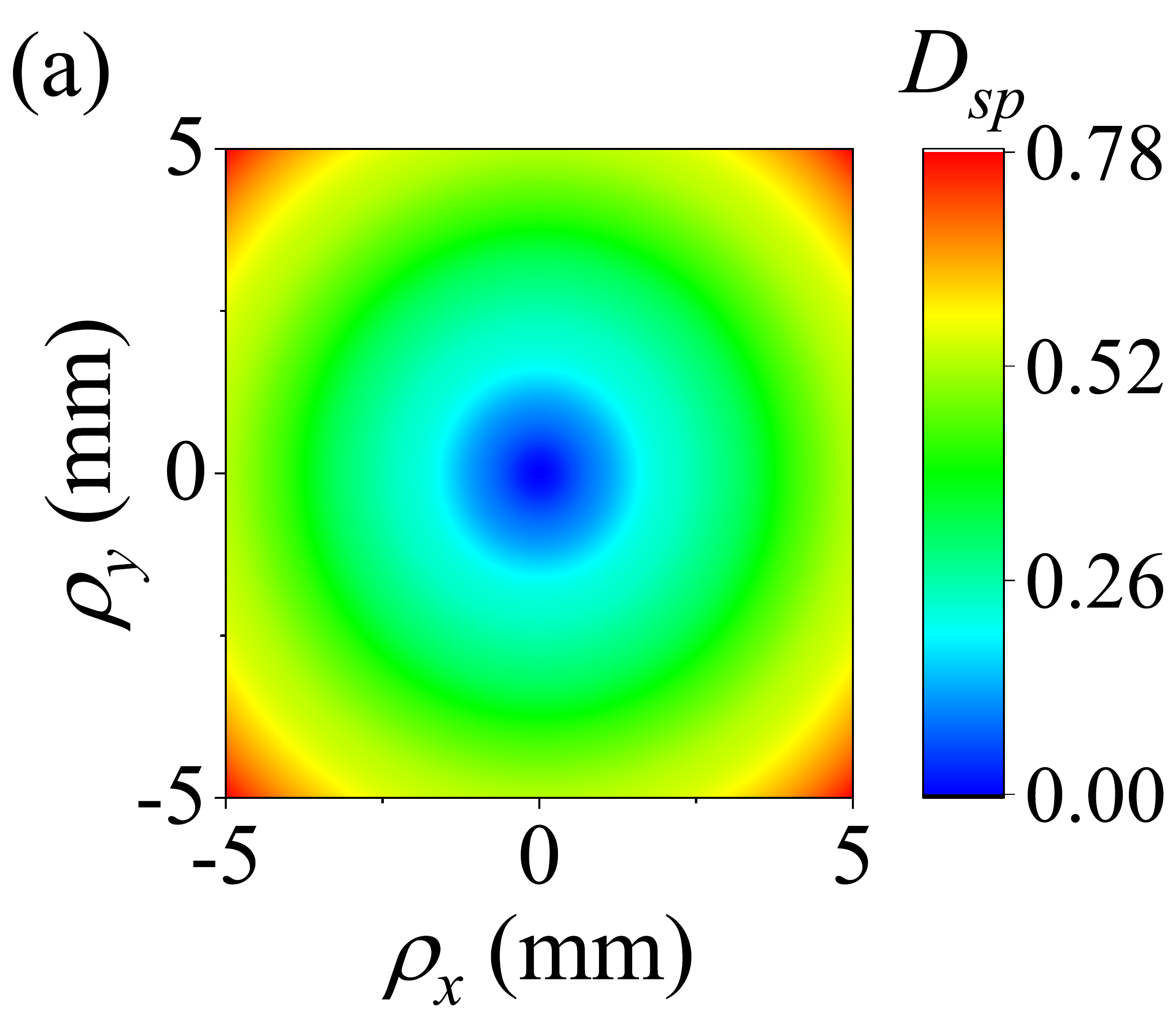}
	\includegraphics[width=0.24\linewidth]{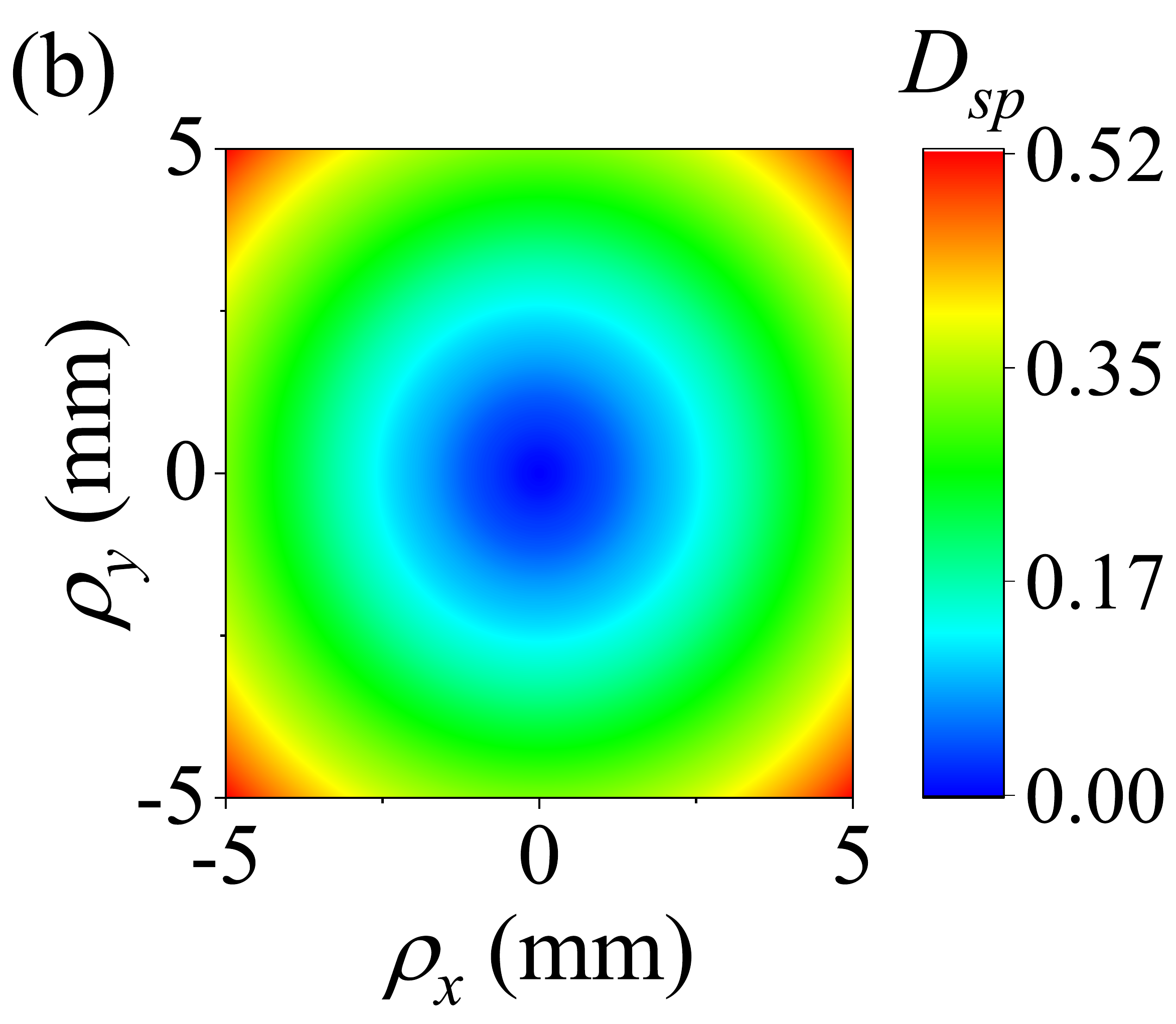}
	\includegraphics[width=0.24\linewidth]{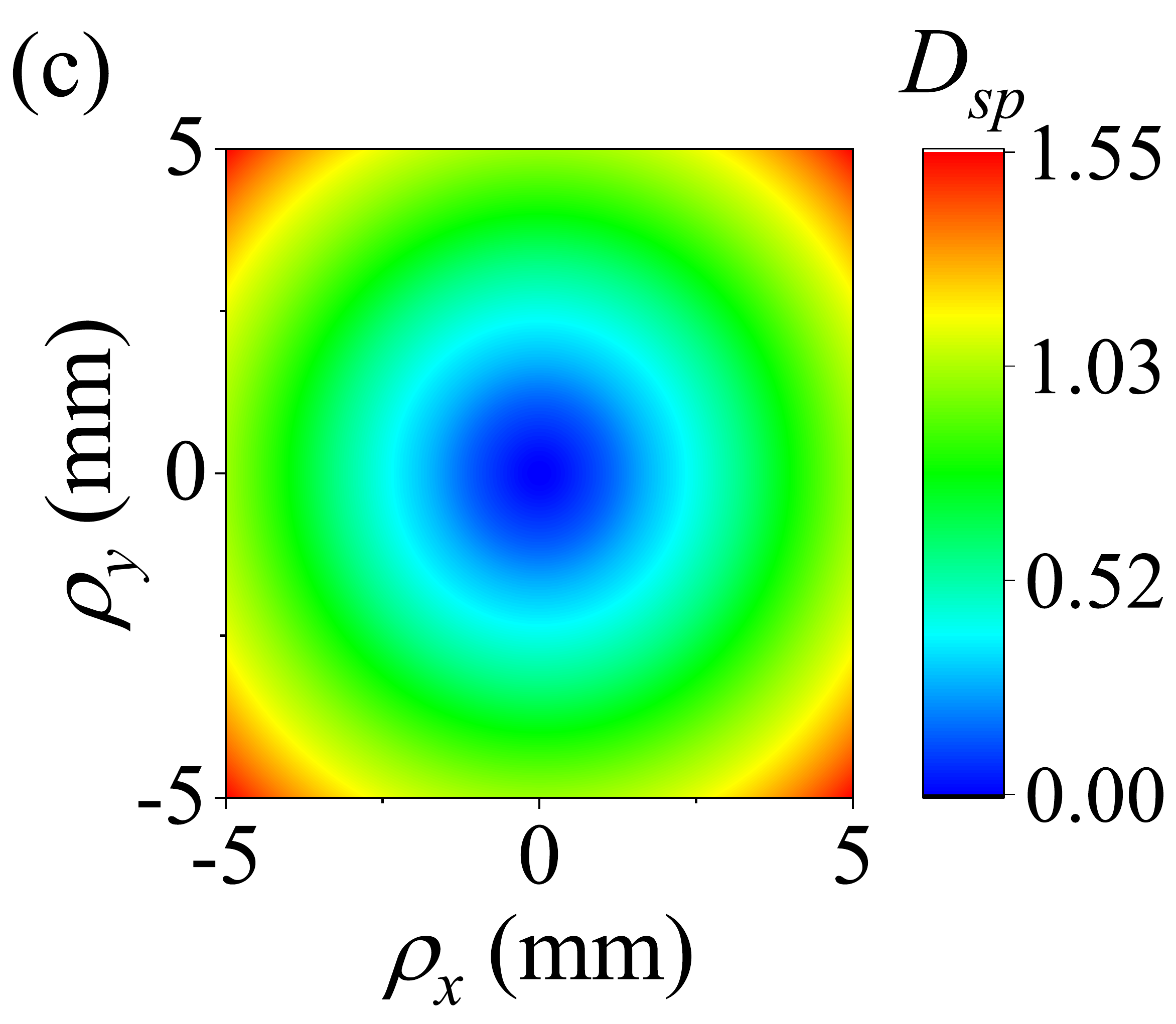}
	\includegraphics[width=0.24\linewidth]{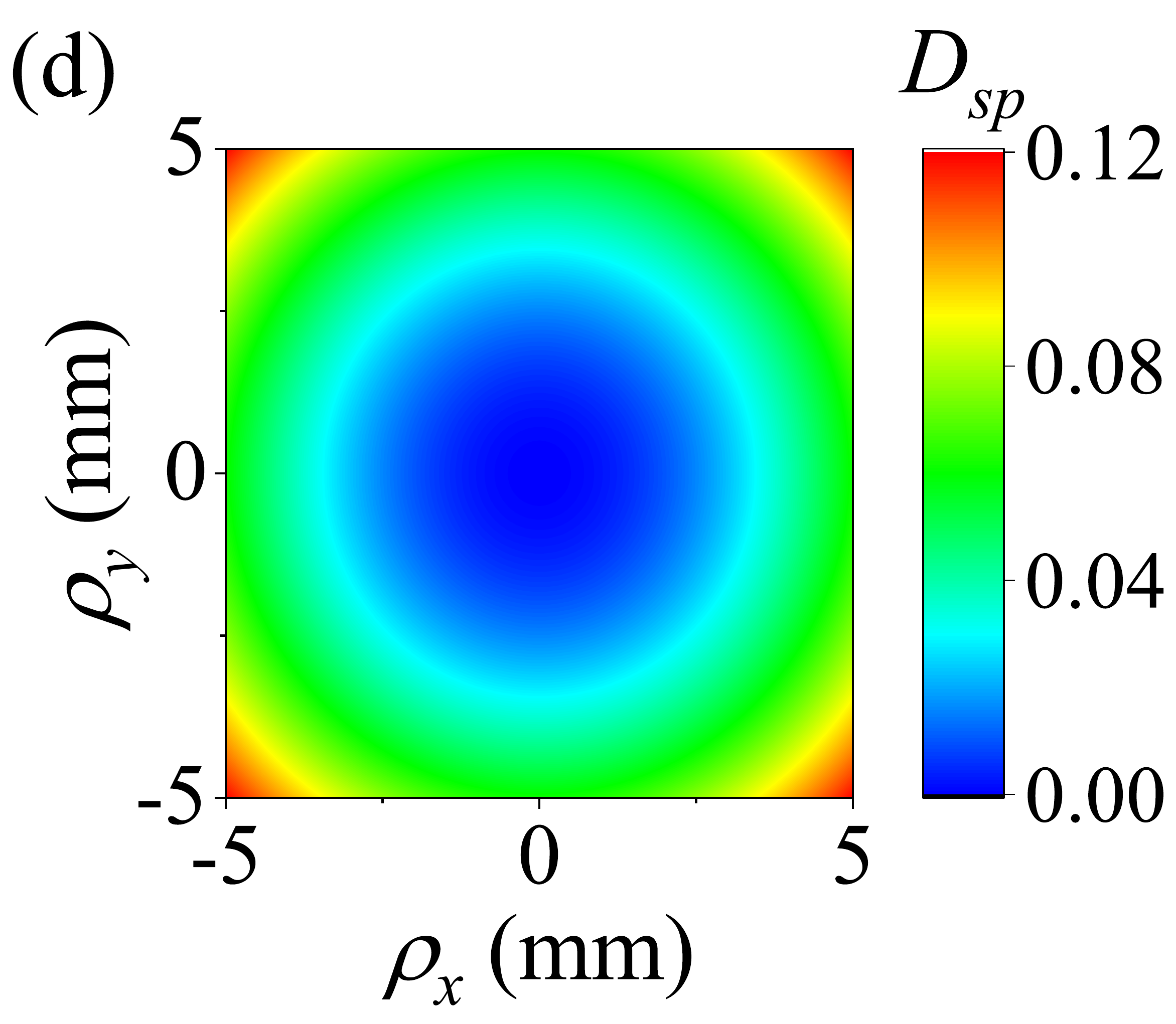}
	\caption{The WSFs of spherical wave in isotropic turbulence ($\mu=1$) with different values of power laws. (a) $\left(\alpha_T,\alpha_S\right) = \left(11/3,11/3\right)$, (b) $\left(\alpha_T,\alpha_S\right) = \left(14/3,11/3\right)$, (c) $\left(\alpha_T,\alpha_S\right) = \left(11/3,14/3\right)$ and (d) $\left(\alpha_T,\alpha_S\right) = \left(14/3,14/3\right)$. Values of other parameters are listed in Appendix II.}
	\label{fig:5}
\end{figure*}

\begin{figure*}[!h]
	\centering
	\includegraphics[width=0.3\linewidth]{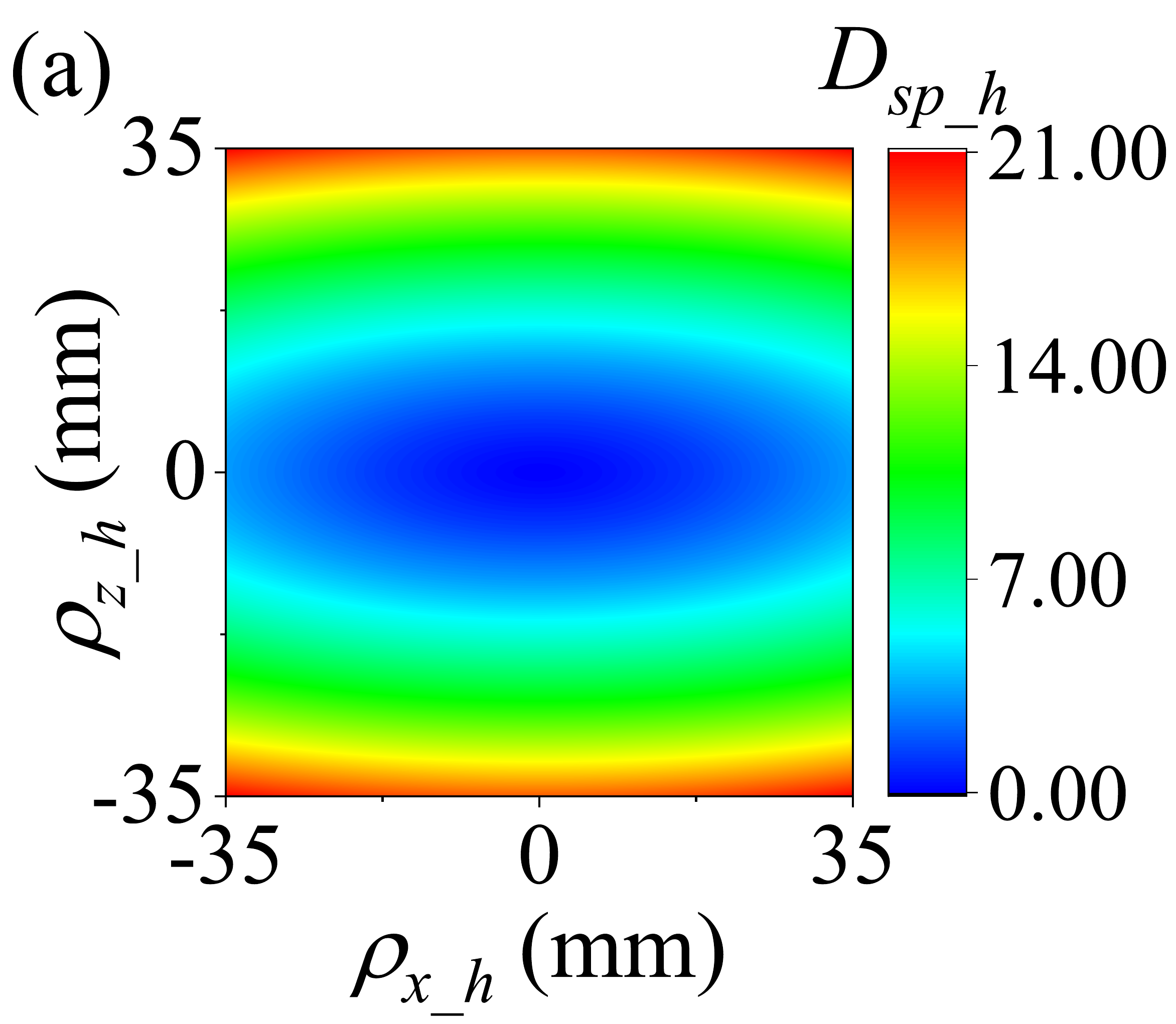}
	\includegraphics[width=0.3\linewidth]{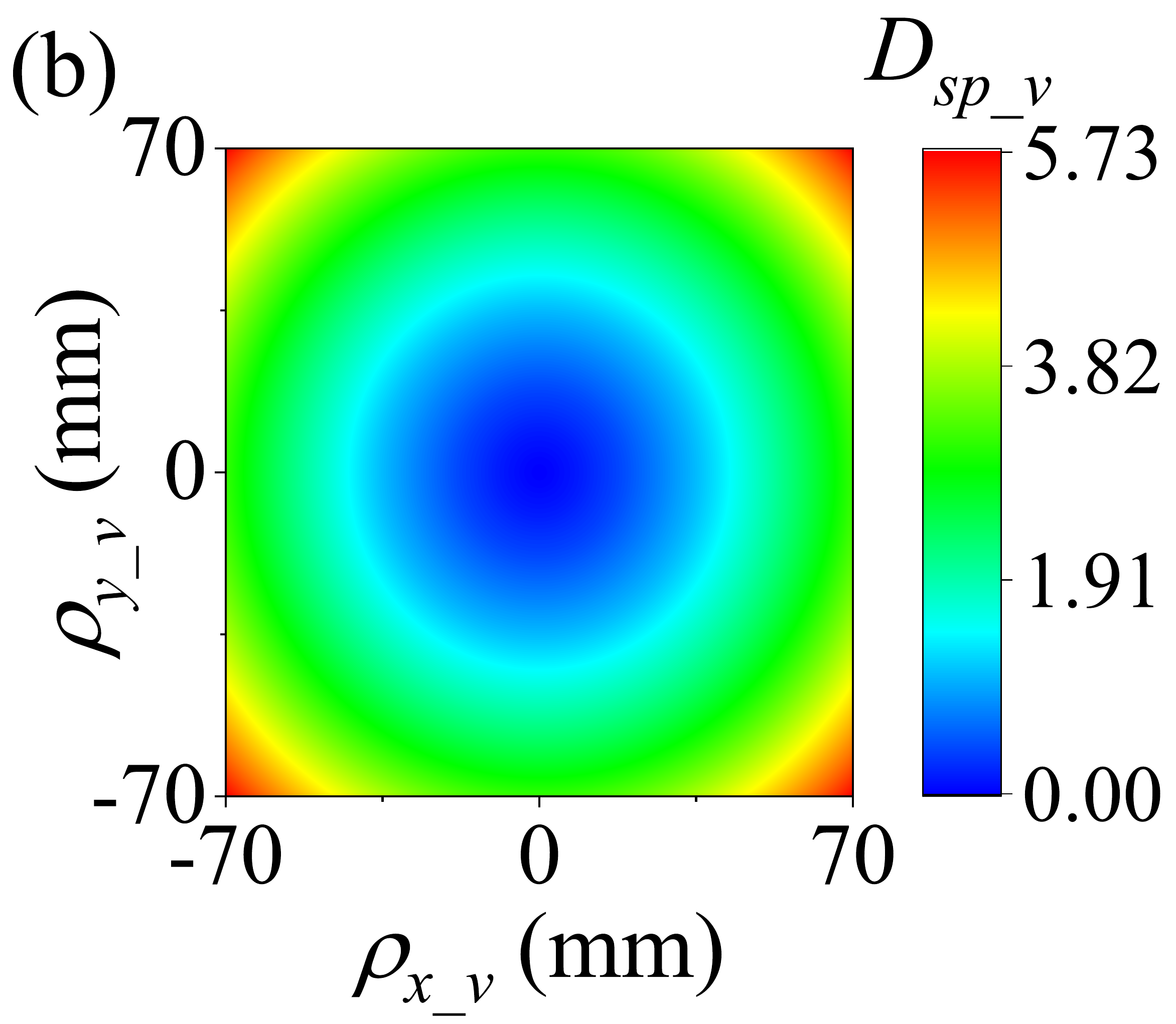}
	\includegraphics[width=0.3\linewidth]{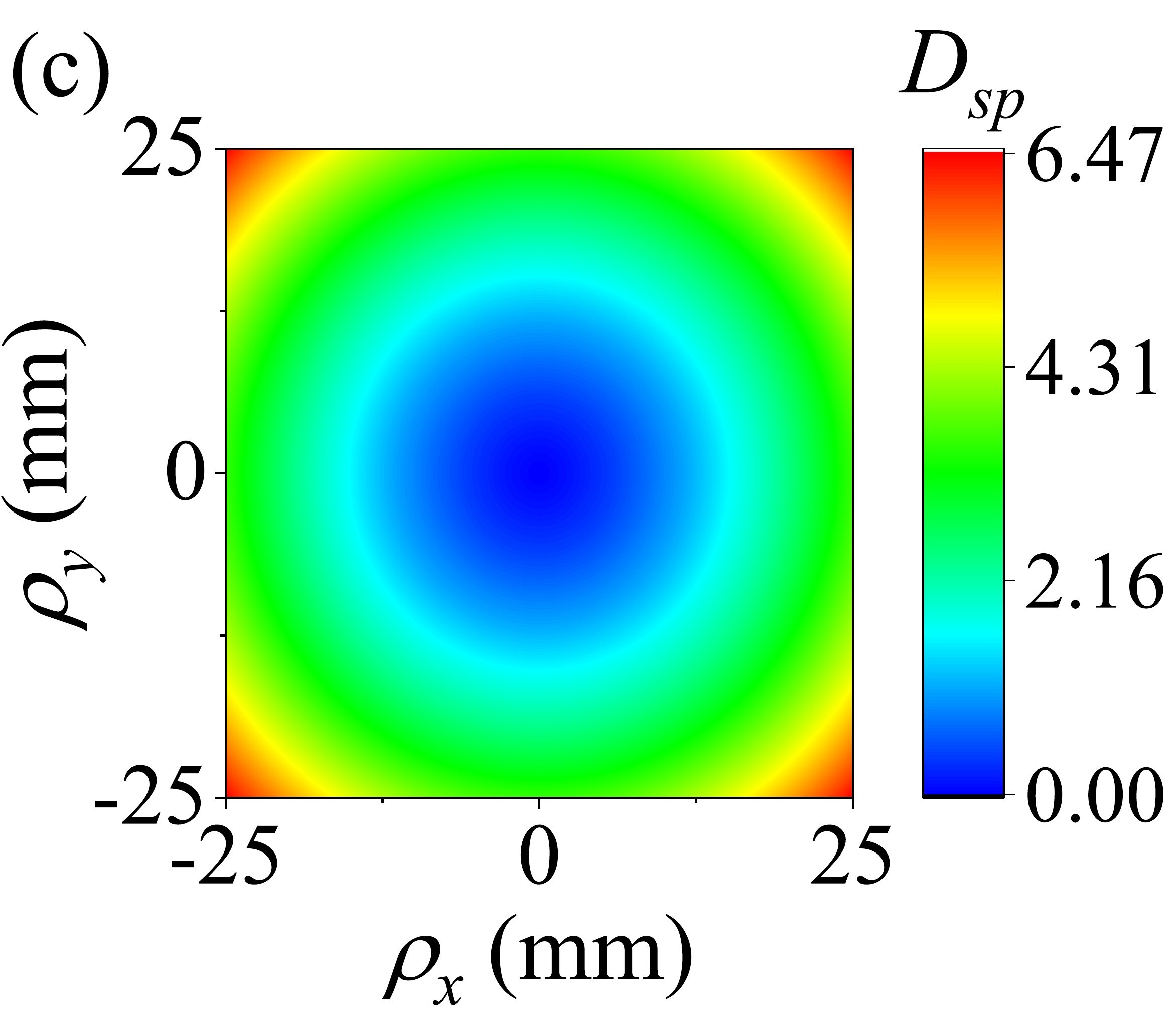}
	\caption{The WSFs of spherical wave (a) in horizontal channels with $\mu = 3$, (b) in vertical channels with $\mu = 3$, and (c) in horizontal/vertical channels  with $\mu = 1$. Values of other parameters are listed in Appendix II.}
	\label{fig:6}
\end{figure*}

\textit{Equations (\ref{eq49}) - (\ref{eq51}) are the main results of this section. They characterize the WSF of a spherical wave in an anisotropic, non-Kolmogorov turbulence by means of single integrals.}
We first plot the numerical results of the WSFs in an isotropic turbulence, with different values of the power law exponents in Fig. \ref{fig:5}, and then compare isotropic and anisotropic cases in Fig. \ref{fig:6}.
Figure \ref{fig:5} shows that the power-law exponents $\alpha_{T}$ and $\alpha_{S}$ have significant effects on the WSF. Such power laws can result in a much higher or lower WSF in the non-Kolmogorov case than that in the Kolmogorov case. Figure \ref{fig:6} shows that anisotropic turbulence leads to an anisotropic WSF
which results in an elliptically shaped coherence radius, which we will further illustrate in the next section. 

\begin{figure*}[!t]
	\centering
	\includegraphics[width=0.3\linewidth]{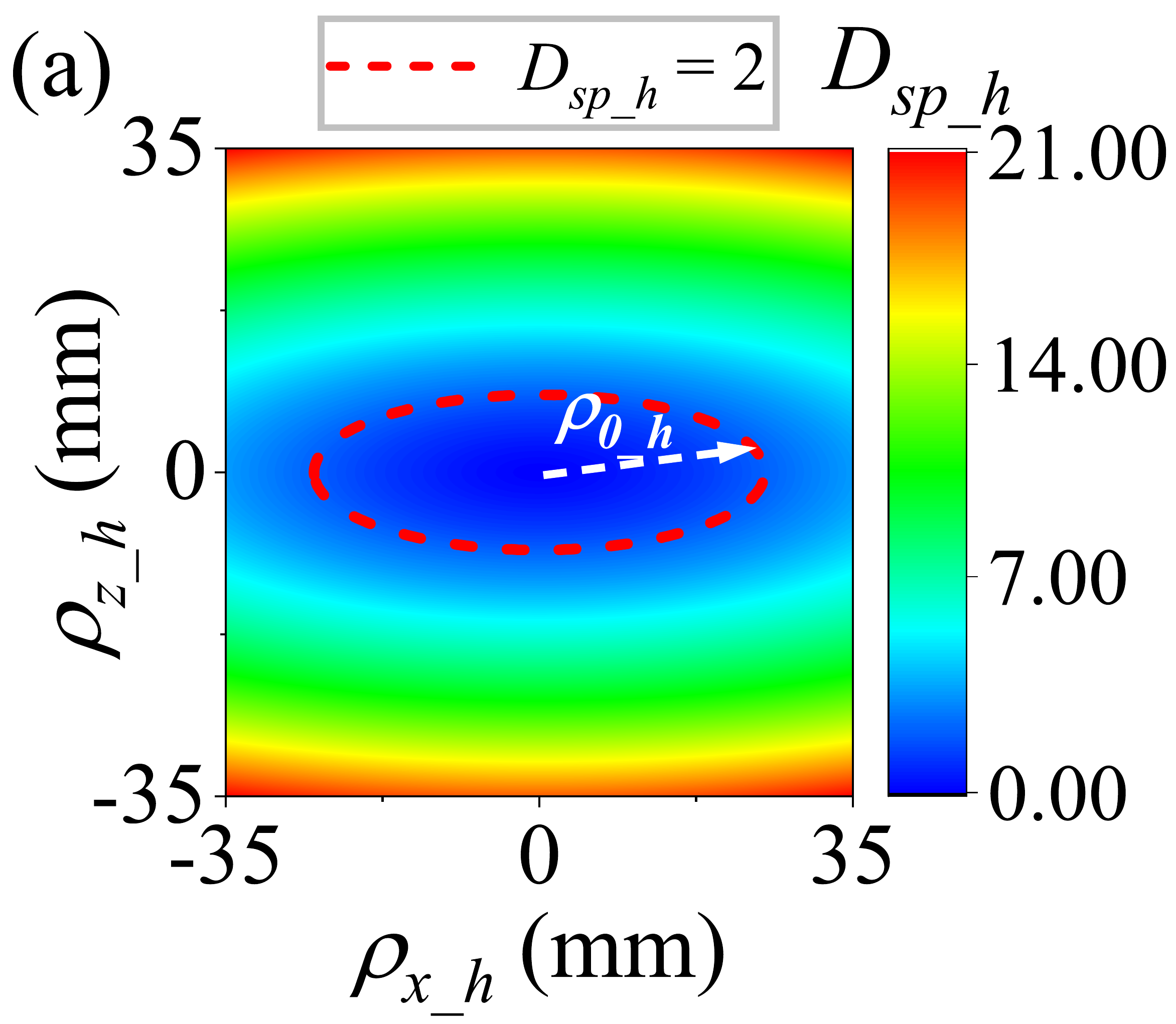}
	\includegraphics[width=0.3\linewidth]{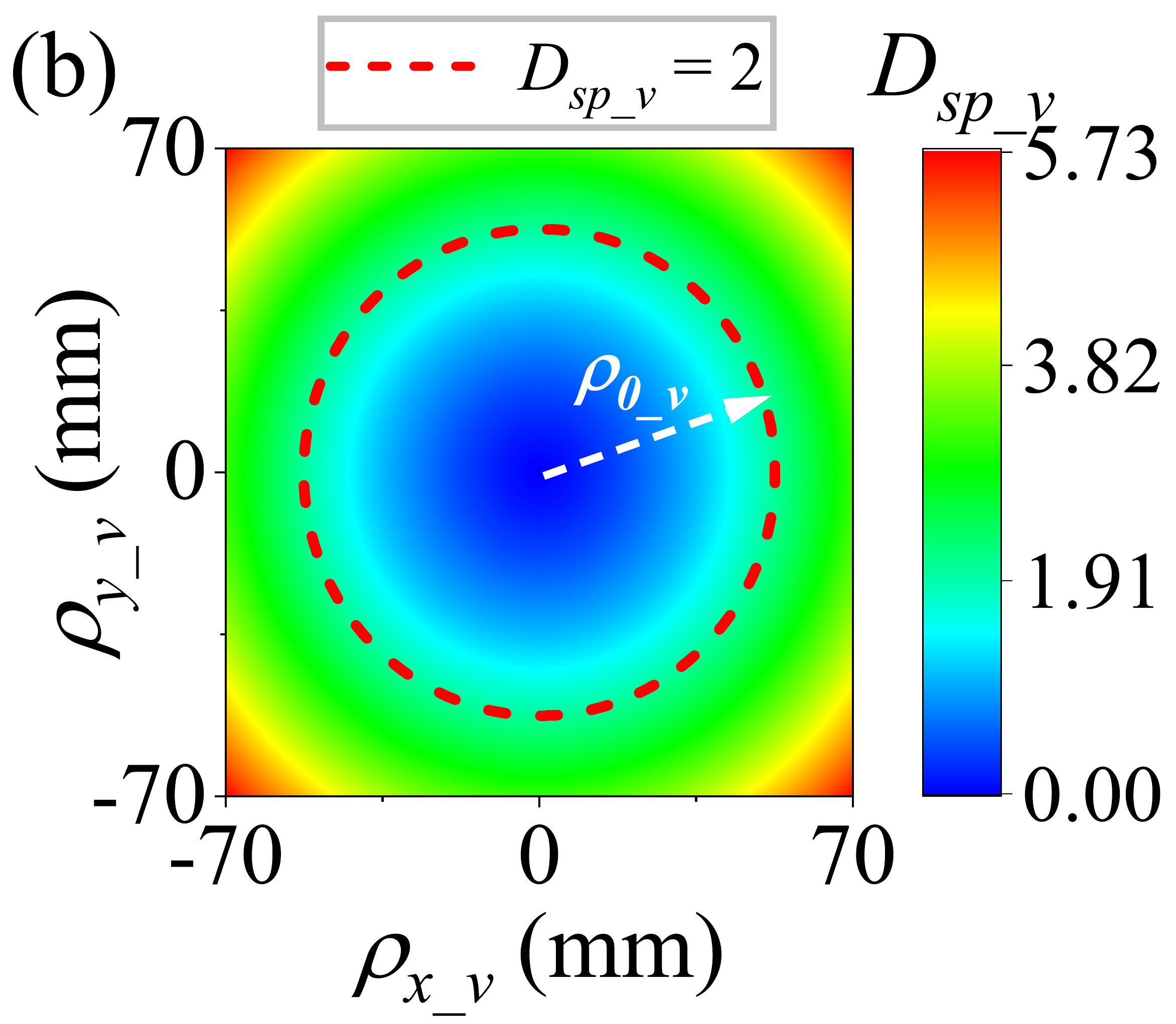}
	\includegraphics[width=0.3\linewidth]{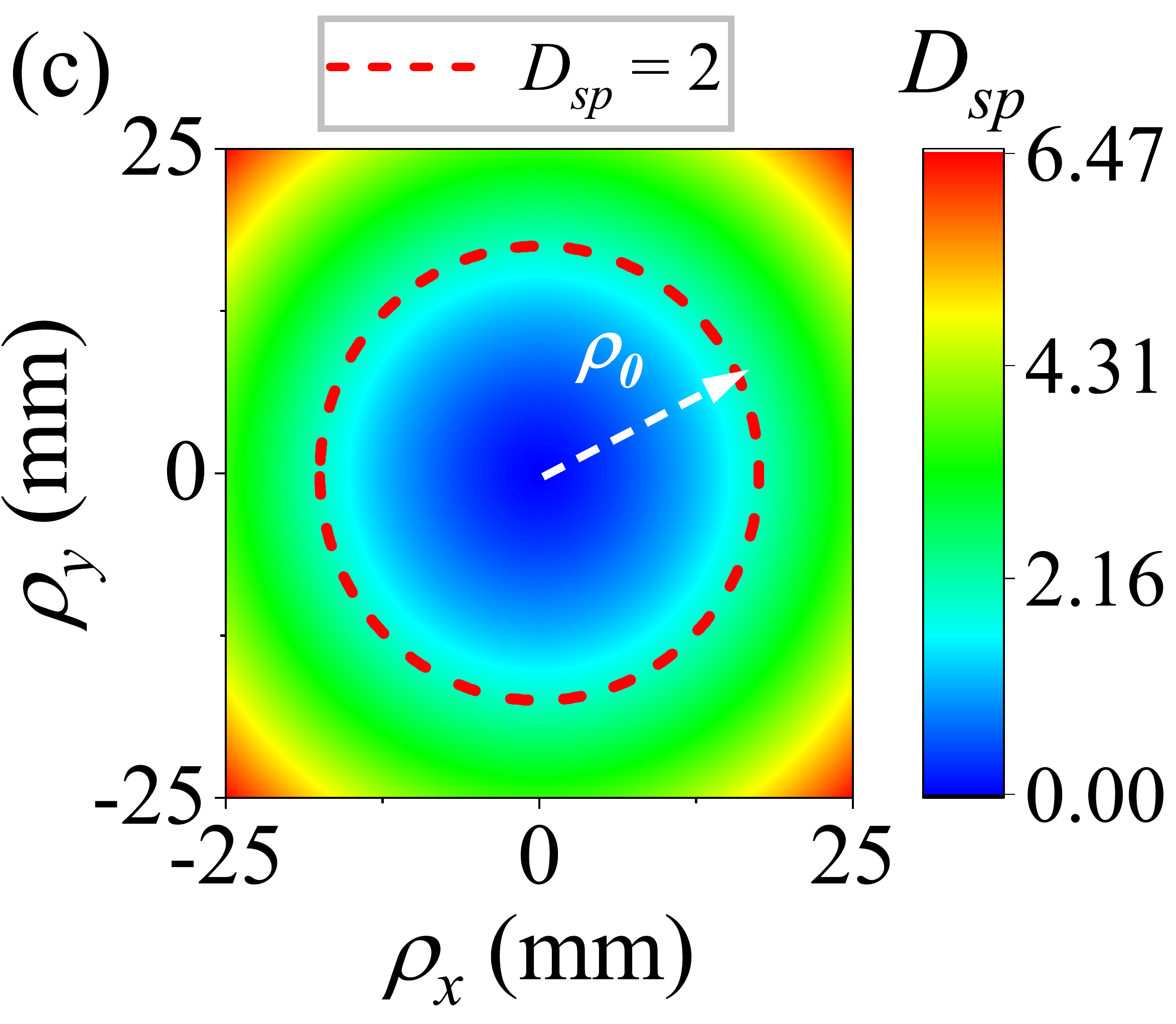}
	\caption{The CRVs of (a) horizontal channel with $\mu = 3$, (b) vertical channel with $\mu = 3$, and (c) horizontal/vertical channel with $\mu = 1$. Values of other parameters are same as those of fig. 6.}
	\label{fig:7}
\end{figure*}

\subsection{Coherence radius of a spherical wave}
The coherence radius of a spherical wave can be directly employed for assessing the optical turbulence strength, and is also useful in calculating the statistics of various optical beams (e.g. \cite{KOROTKOVA2020}). It is defined as a transverse separation distance between two points in the propagating spherical wave that corresponds to the WSF's value of 2. As a rule, the coherence radius is considered to be a scalar quantity. 

However, as we have shown in Section 4.1, the WSF could be anisotropic. Hence, here the `coherence radius' is considered as a vector $\bm{\rho}_0$ and we define it as a \textit{coherence radius vector} (CRV) $\bm{\rho_0}$ by setting 
\begin{equation}
{D_{sp}}\left( {{\bm{\rho}_0},L} \right) = 2.
\label{eq52}
\end{equation}
For horizontal and vertical channels, we rewrite Eq. (\ref{eq52}) as
\begin{equation}
{D_{sp\_h}}\left( {{\bm{\rho} _{0\_h}},L} \right) = 2,\ {D_{sp\_v}}\left( {{\bm{\rho} _{0\_v}},L} \right) = 2,
\label{eq53}
\end{equation}
where ${\bm{\rho} _{0\_h}} = {\left( {{\rho _{0x\_h}},{\rho _{0z\_h}}} \right)^T}$, ${\bm{\rho} _{0\_v}} = {\left( {{\rho _{0x\_v}},{\rho _{0y\_v}}} \right)^T}$ are the CRVs in horizontal channel and vertical channel, respectively. 

A \textit{coherence radius scalar} (CRS) $\rho _{0\_{\rm{iso}}}$ is assumed as
\begin{equation}
{\rho _{0\_{\rm{iso}}}} = \left\{ {\begin{array}{*{20}{c}}
	{\sqrt {\rho _{0x\_h}^2 + {\mu ^{2}}\rho _{0z\_h}^2} }&{{\rm{in}}\;{\rm{horizontal}}\;{\rm{channel}},}\\
	{\sqrt {\rho _{0x\_v}^2 + \rho _{0y\_v}^2} }&{{\rm{in}}\;{\rm{vertical}}\;{\rm{channel}}.}
	\end{array}} \right.
\label{eq54}
\end{equation}
$\rho _{0\_{\rm{iso}}}$ equals the widely used coherence radius if $\mu = 1$ or in vertical channel. Combining Eqs. (\ref{eq49})-(\ref{eq50}), (\ref{eq53}) and (\ref{eq54}), we have
\begin{equation}
\frac{{8{\pi ^2}{k^2}L}}{{n_0^2}}\int_0^{ + \infty } {d\kappa  \cdot \kappa {\Phi _{n1}}(\kappa )\left[ {1 - {J_0}\left( {\mu^{-1}\kappa {\rho _{0\_{\rm{iso}}}}} \right)} \right]}  = \left\{ {\begin{array}{*{20}{c}}
	{2 \mu^{-1}}&{{\rm{for}}\;{\rm{horizontal}}\;{\rm{channels}},}\\
	2&{{\rm{for}}\;{\rm{vertical}}\;{\rm{channels}},}
	\end{array}} \right.
\label{eq55}
\end{equation}
where $\Phi _{n1}$ is the outer-scaled NK-OTOPS. Eqs. (\ref{eq54}) and (\ref{eq55}) can be used to predict the CRS $\rho_{0\_{\rm{iso}}}$ and the CRV $\bm{\rho_0}$ in oceanic turbulence. For example, according to Eqs. (\ref{eq38}) and (\ref{eq55}), $\rho_{0\_{\rm{iso}}}$ in the cases of Figs. \ref{fig:6} (a)-(c) are $25.1 \rm{mm}$, $52.5 \rm{mm}$ and $17.5 \rm{mm}$, respectively; substituting $\rho_{0\_{\rm{iso}}}$ into Eq.(\ref{eq54}), we mark the CRVs by white arrows in Fig. \ref{fig:7}.

\textit{The derived coherence radius vector (CRV) and scalar (CRS) are the main results of this section, which can be evaluated using Eqs. (\ref{eq54})-(\ref{eq55}).} The CRS corresponds to the widely used coherence radius if $\mu = 1$ or along a vertical channel, and it could measure the anisotropic turbulence strength along different directions. In fact, the atmospheric turbulence anisotropy has been recently directly assessed through a measurement of the elliptically shaped mutual coherence function of a laser beam \cite{Davis} (see also a similar measurement via the elliptically shaped intensity-intensity correlation function \cite{AnisEllipse}).   

\subsection{Co-effect of temperature and salinity on coherence radius scalar}
In this section we will give a numerical example on the co-effect of temperature and salinity of the NK-OTOPS on the CRS by calculating it as a function of the power laws of temperature and salinity spectra $\alpha_{T}$, $\alpha_{S}$, as well as parameters $c_T$ and $c_S$, defined by Eq. (\ref{eq6}).

For brevity of discussion, we set the anisotropy constant $\mu = 3$ \cite{footnote6}, and choose the CRS $\rho_{0\_{\rm{iso}}}$ in vertical channels as a measurement of turbulent disturbance.
The ranges of related parameters are listed as follows (see more details in Appendix. I):
\begin{center}
\fbox{
	\shortstack[l]{$\cdot\,$$\alpha_{T}, \alpha_{S} \in \left[11/3, 15/3\right)$;\\
$\cdot\,$${c_T} \in \left[ {1.61\times 10^{-3},3.99\times 10^{-3}} \right]$ and ${c_S} \in \left[ {9.76\times 10^{-6},61.62\times10^{-6}} \right]$;\\
$\cdot\,$${C_S^2}/{C_T^2} \ge 3.18\times10^{-5}\rm{ppt}^2\cdot\rm{deg}^{-2}\cdot\rm{m}^{\alpha_{T}-\alpha_{S}}$.}}
\end{center}

\begin{figure}[!t]
	\centering
	\subfigure[Full correlation:$\gamma_{ST} = 1$;]{
		\centering
		\includegraphics[width=0.23\linewidth]{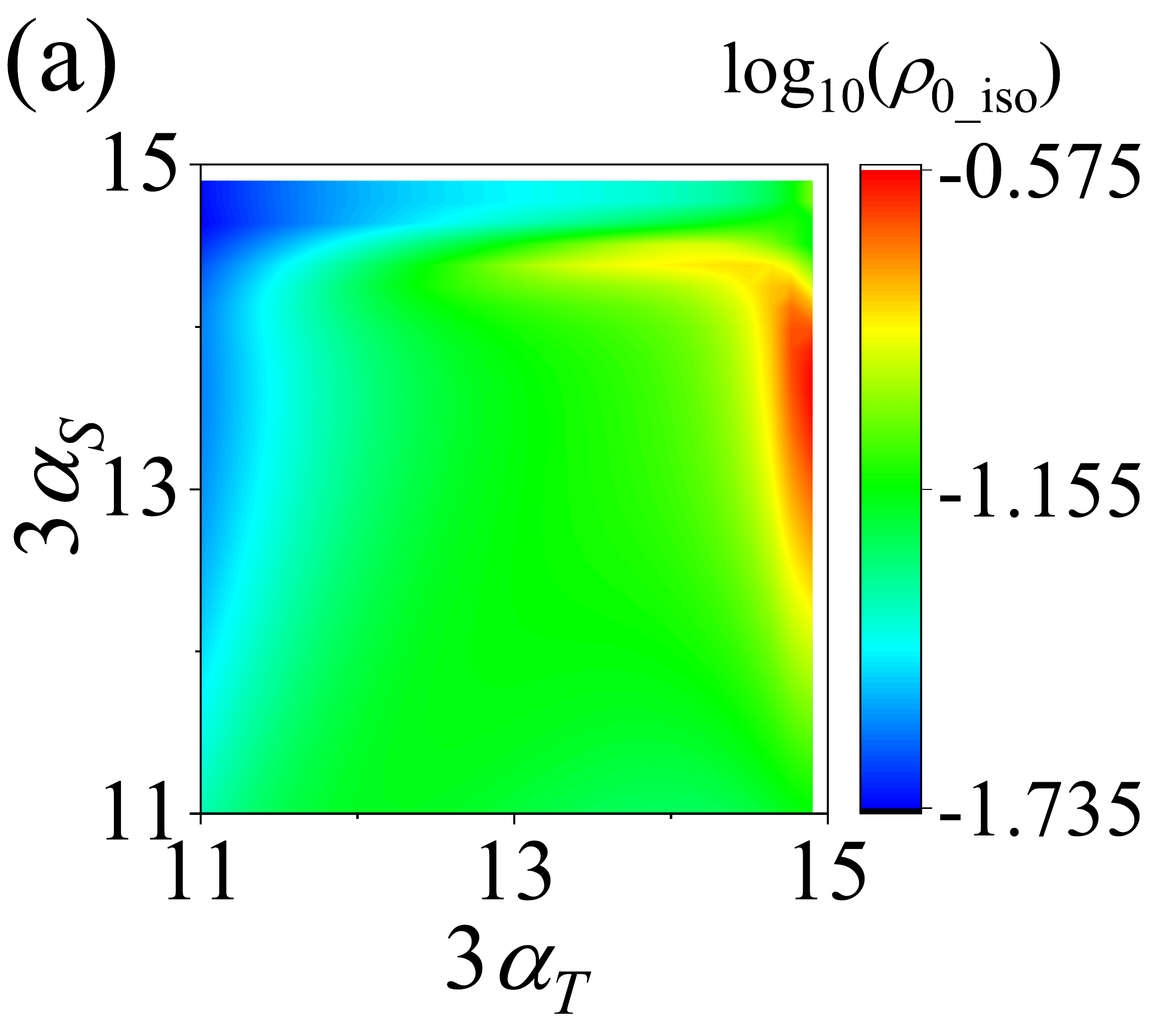}	
	}
	\subfigure[Partial correlation:$p = 2$;]{
		\centering
		\includegraphics[width=0.23\linewidth]{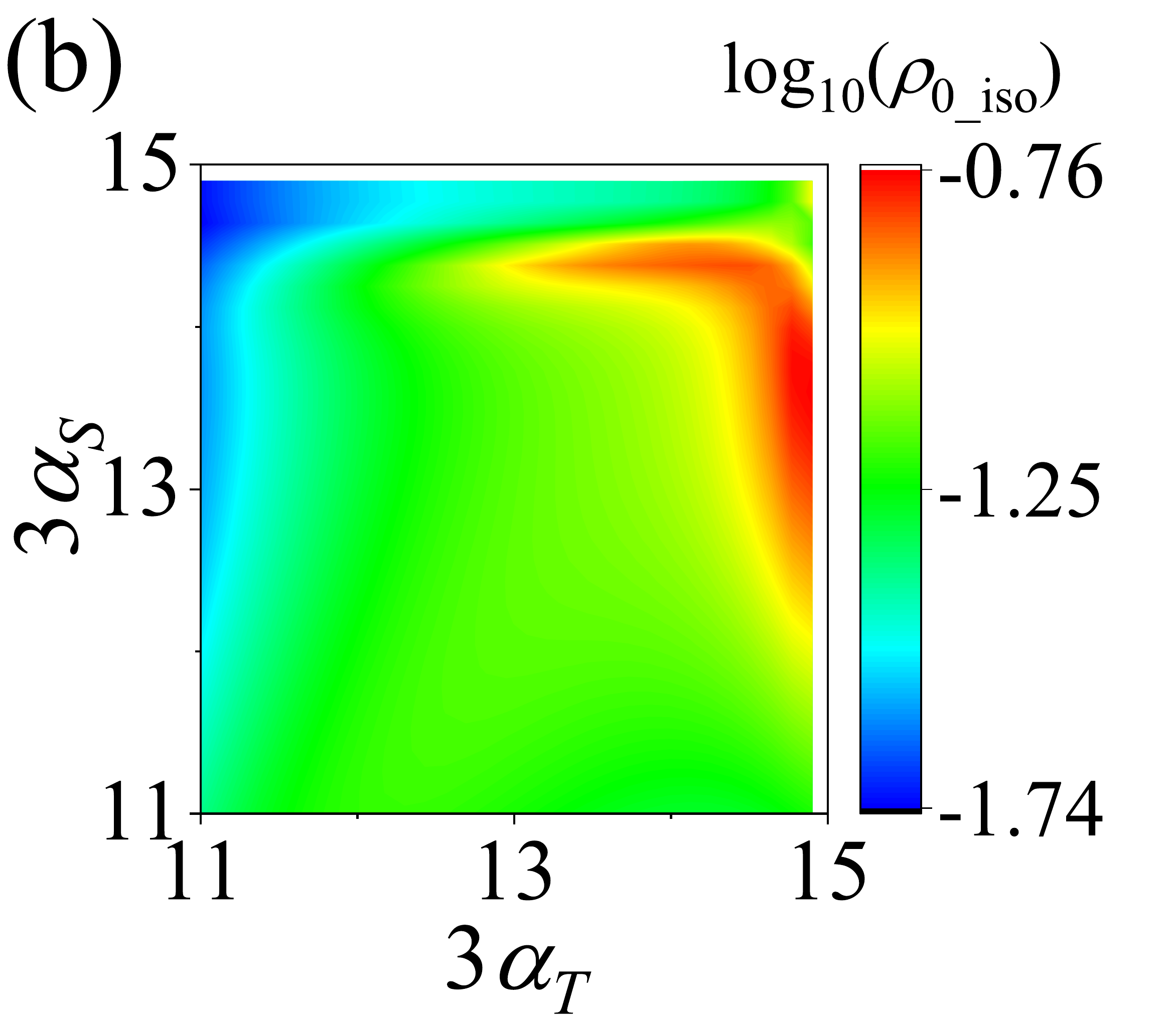}	
	}
	\subfigure[Partial correlation:$p = 4$;]{
		\centering
		\includegraphics[width=0.23\linewidth]{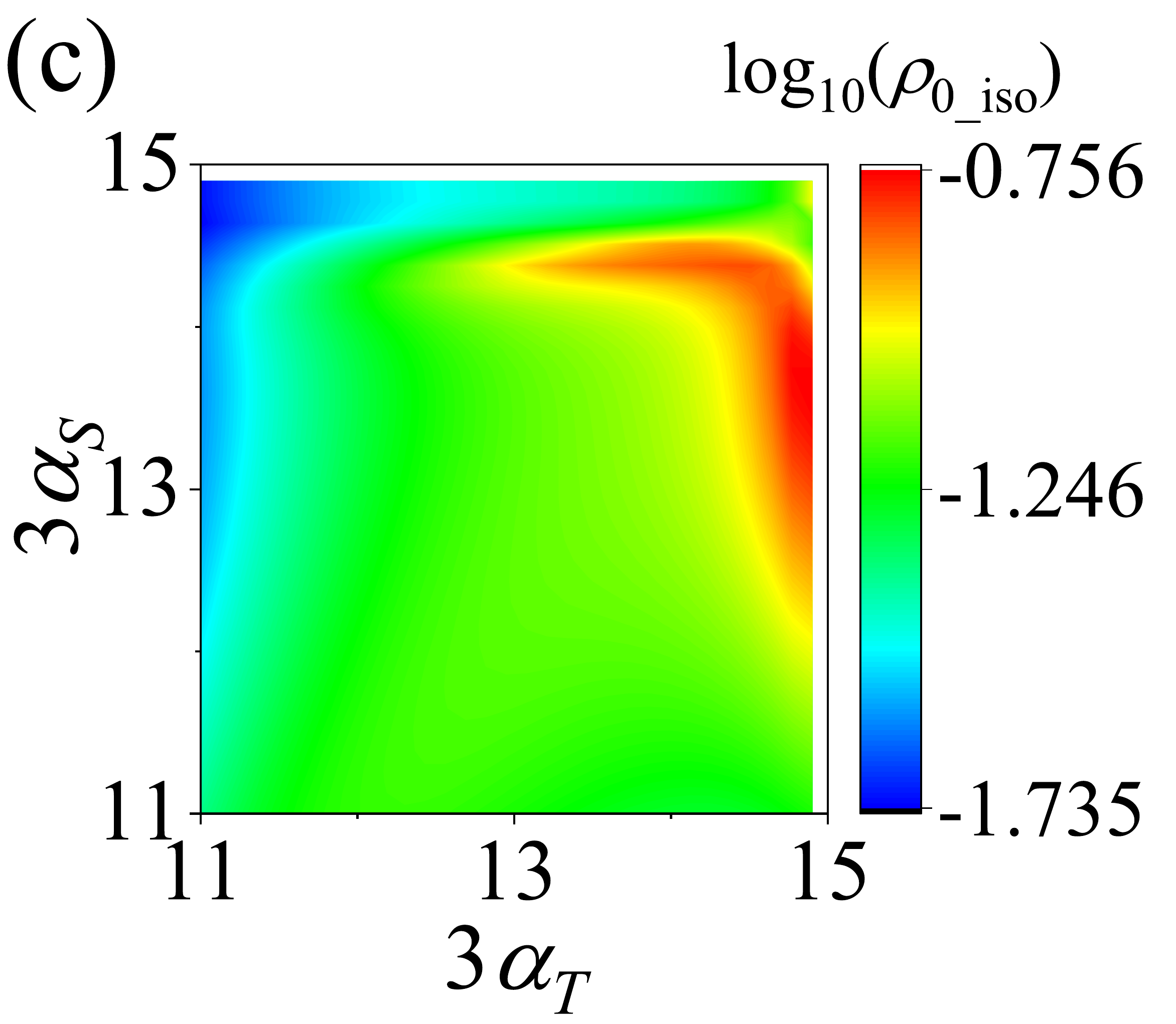}	
	}
	\subfigure[Non-correlation:$\gamma_{ST} = 0$;]{
		\centering
		\includegraphics[width=0.24\linewidth]{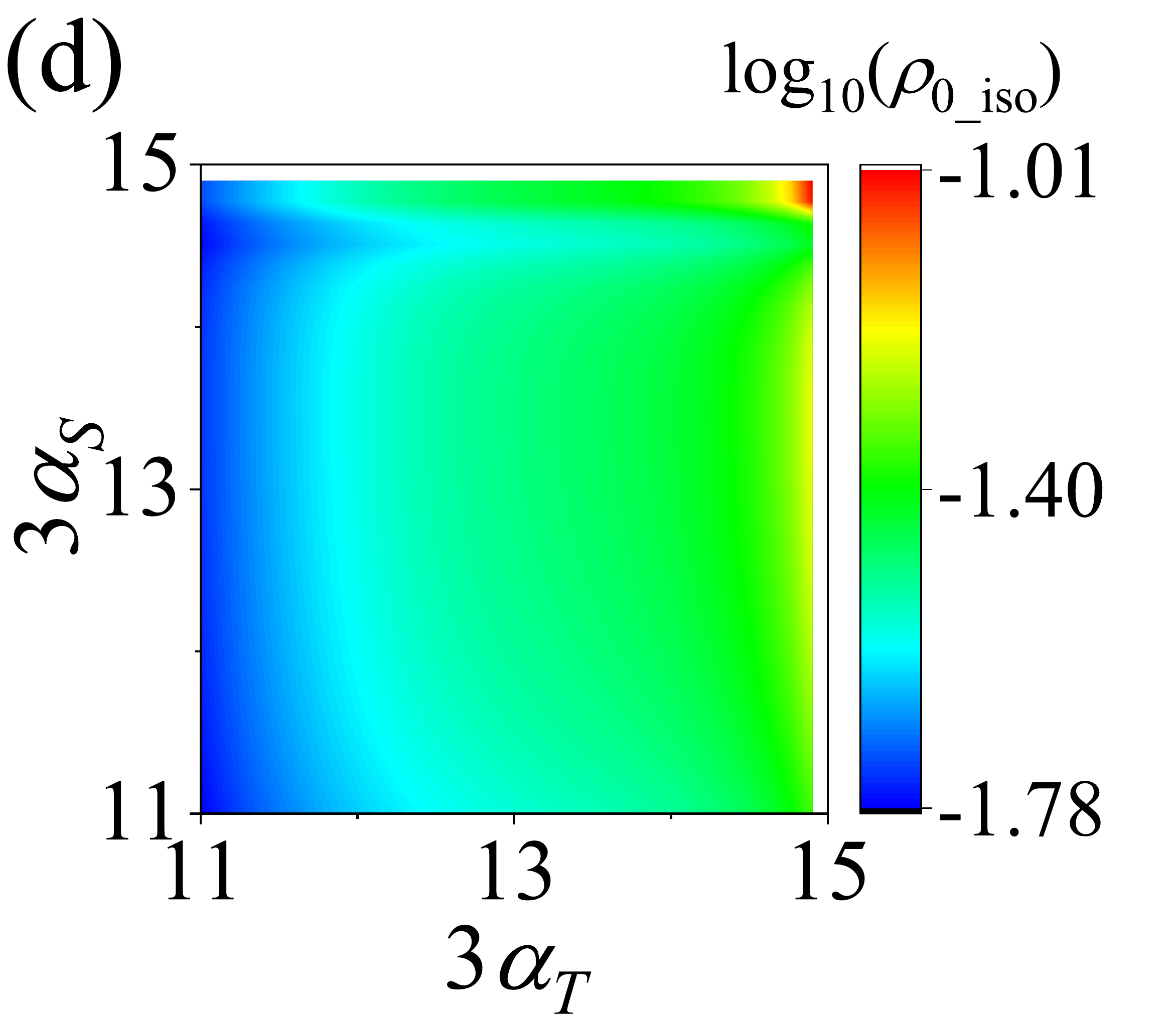}	
	}
	\\
	\subfigure[Full correlation:$\gamma_{ST} = 1$;]{
		\centering
		\includegraphics[width=0.23\linewidth]{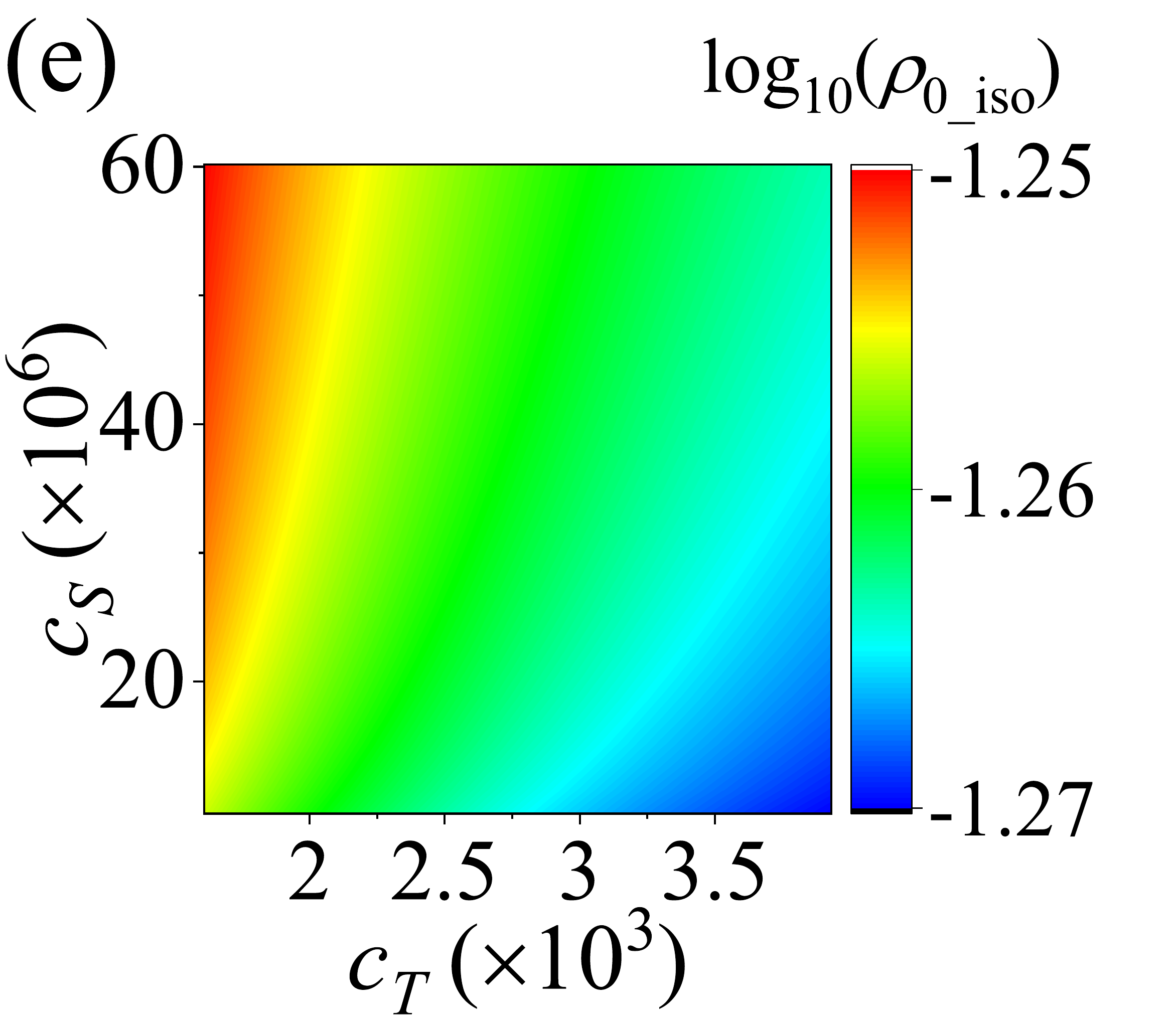}	
	}
	\subfigure[Partial correlation:$p = 2$;]{
		\centering
		\includegraphics[width=0.23\linewidth]{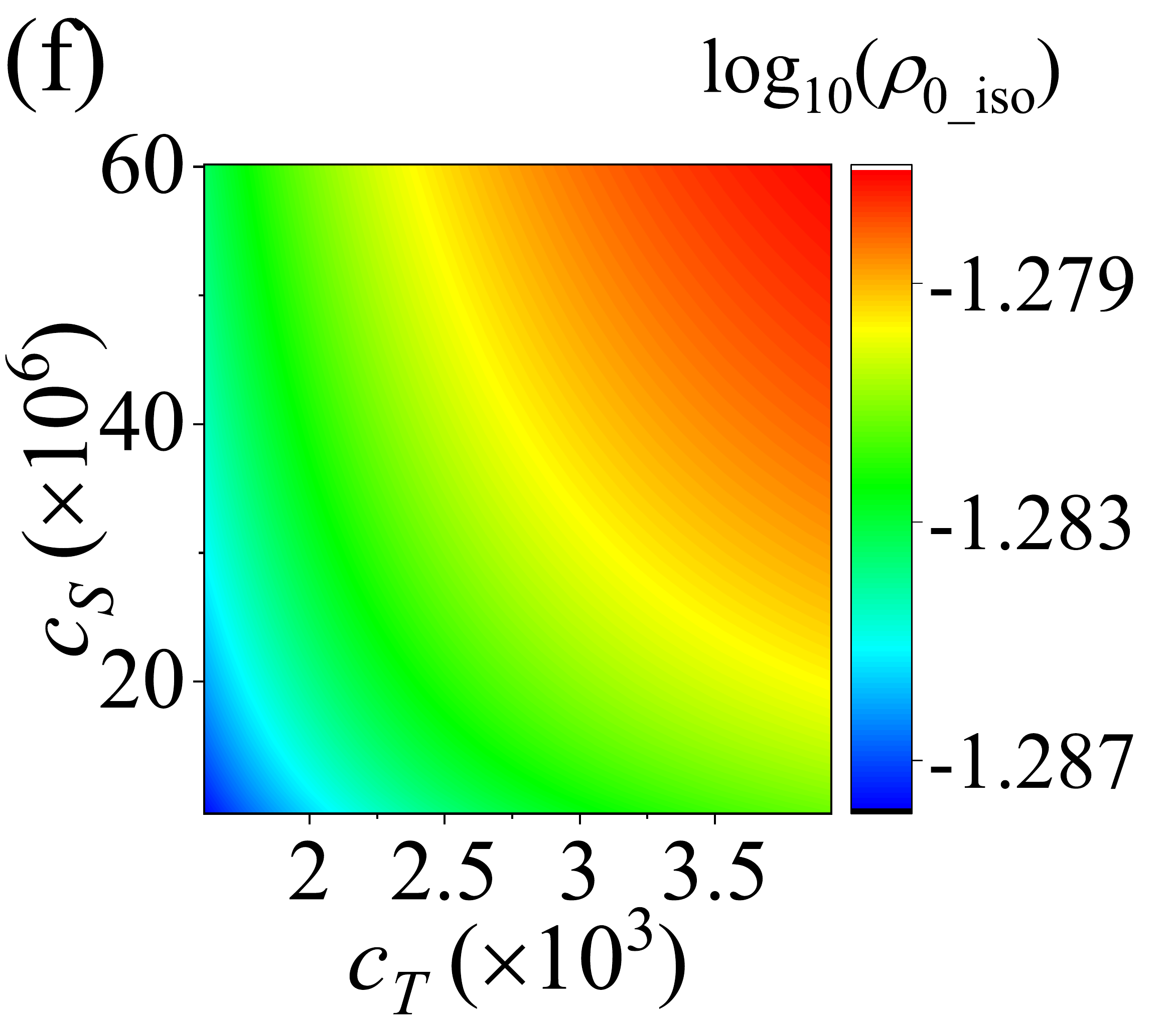}	
	}
	\subfigure[Partial correlation:$p = 4$;]{
		\centering
		\includegraphics[width=0.23\linewidth]{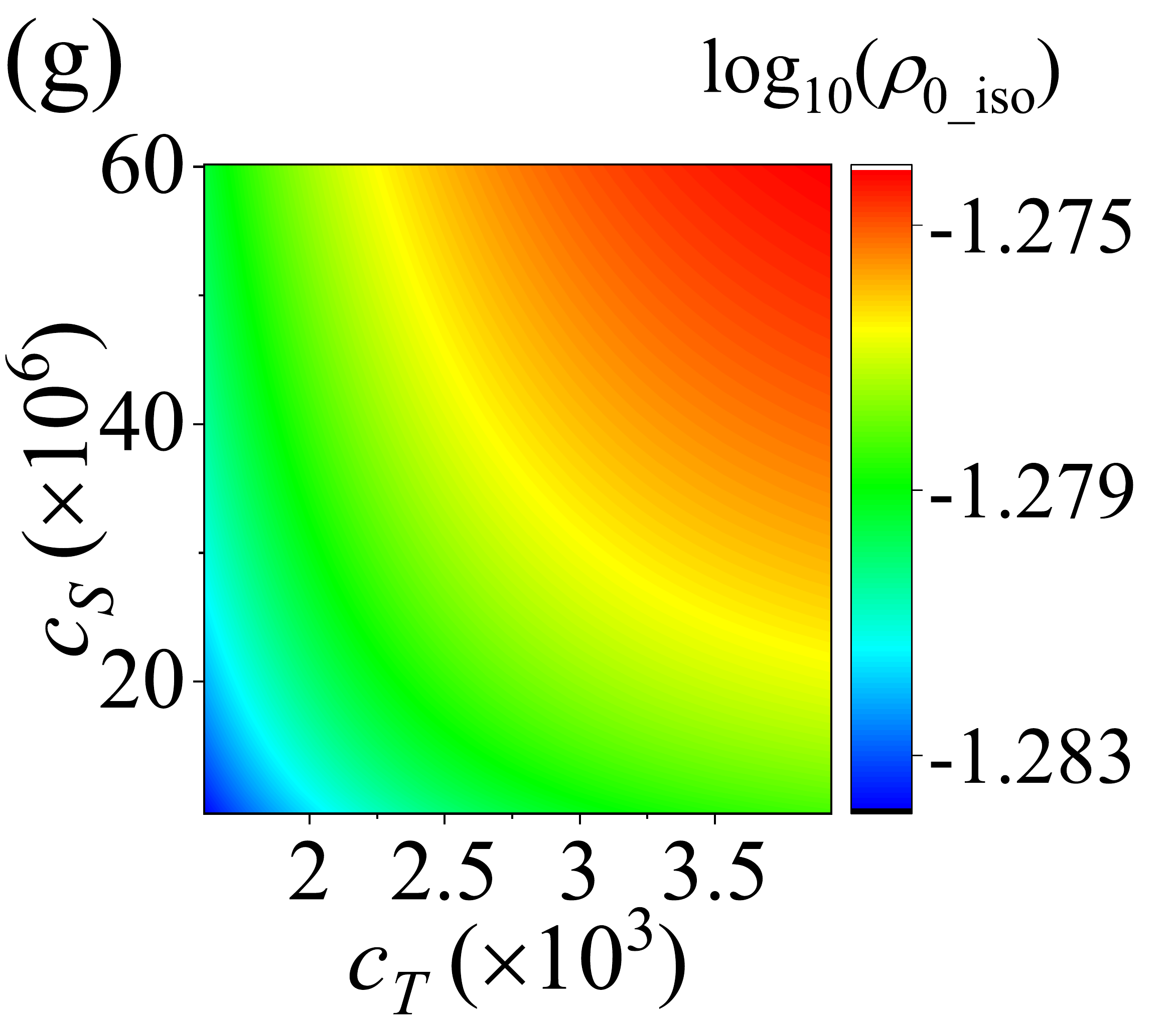}	
	}
	\subfigure[Non-correlation:$\gamma_{ST} = 0$;]{
		\centering
		\includegraphics[width=0.24\linewidth]{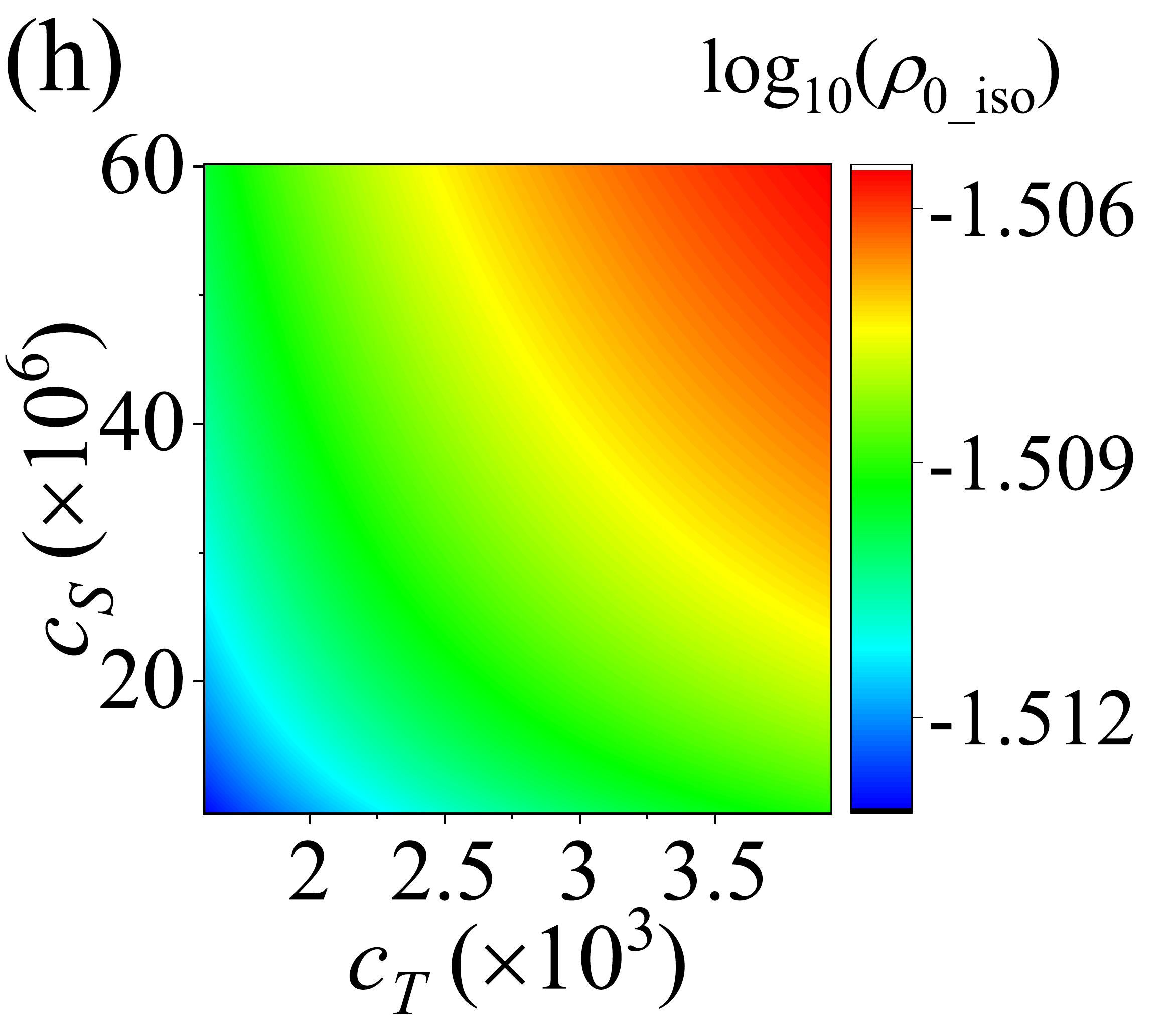}	
	}
	\caption{The distributions $\rho_{0\_{\rm{iso}}}\left(\alpha_{T}, \alpha_{S}\right)$ and $\rho_{0\_{\rm{iso}}}\left(c_T, c_S\right)$ with different spectral correlation $\gamma_{ST}$. Values of parameters are listed in Appendix II. Here $\alpha_T$, $\alpha_S$, $c_T$ and $c_S$ are dimensionless.}
	\label{fig:8}
\end{figure}

Figure \ref{fig:8} shows $\rho_{0\_{\rm{iso}}}\left(\alpha_{T}, \alpha_{S}\right)$ and $\rho_{0\_{\rm{iso}}}\left(c_T, c_S\right)$ for different spectral correlation of the power spectrum (as above, fully correlated case refers to $\gamma_{ST}=1$, partially correlated case refers to the $\gamma_{ST}$ obeying Eq. (\ref{eq31}),  and uncorrelated case refers to $\gamma_{ST}=0$, i.e. ${\Phi _{TS}}=0$.). 
Figure \ref{fig:8} (d) shows a distribution of $\rho_{0\_{\rm{iso}}}\left(\alpha_{T}, \alpha_{S}\right)$ being very different from that in Figs. \ref{fig:8} (a)-(c),
and Fig. \ref{fig:8} (e) shows a distribution of $\rho_{0\_{\rm{iso}}}\left(c_T, c_S\right)$ being very different  from that in Figs. \ref{fig:8} (f)-(h).

\begin{figure}
	\centering
	\subfigure[$C^2_S/C^2_T = 3.18\times10^{-5}$;]{
		\centering
		\includegraphics[width=0.26\linewidth]{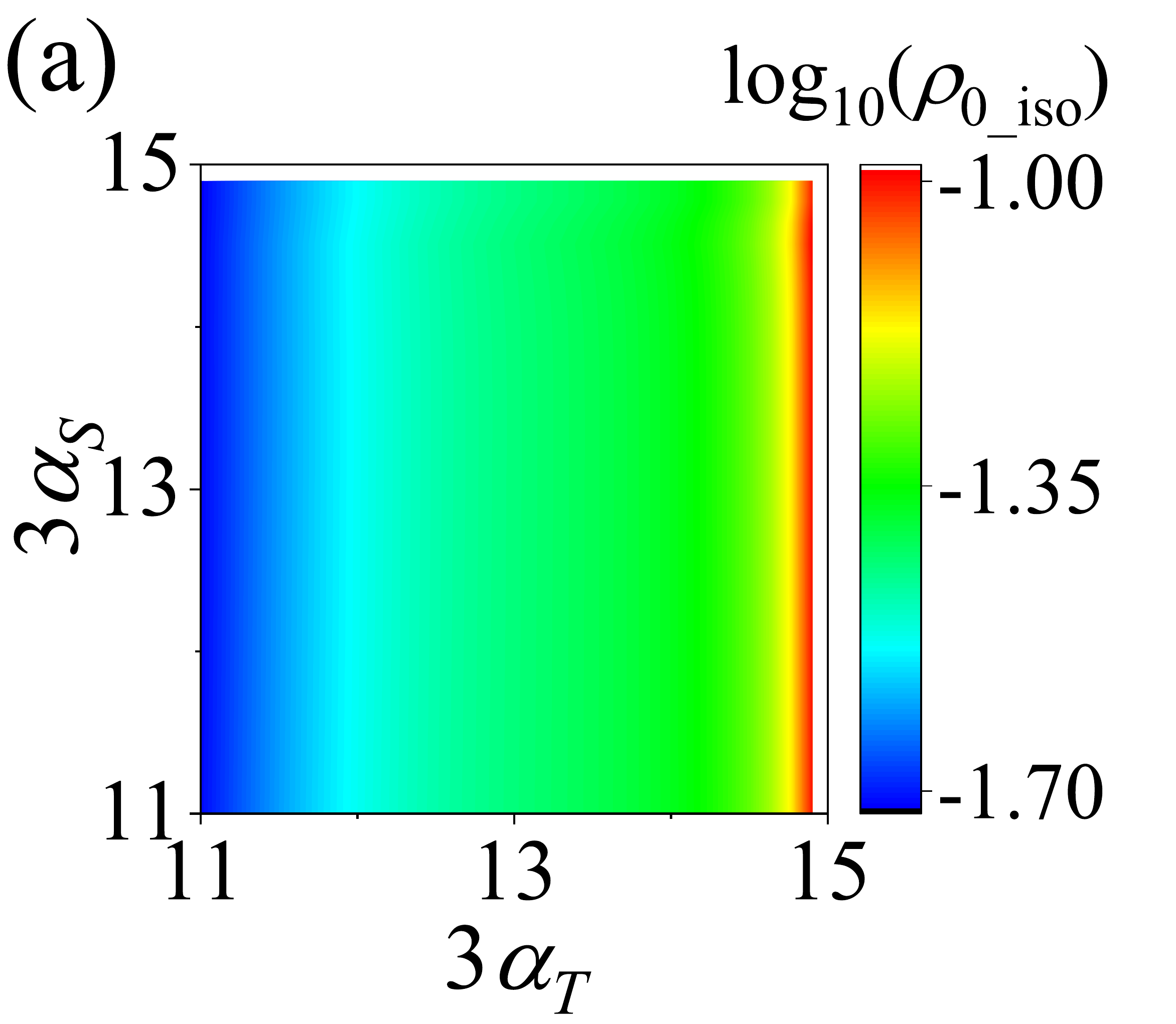}	
	}
	\subfigure[$C^2_S/C^2_T = 3.18\times10^{-3}$;]{
		\centering
		\includegraphics[width=0.26\linewidth]{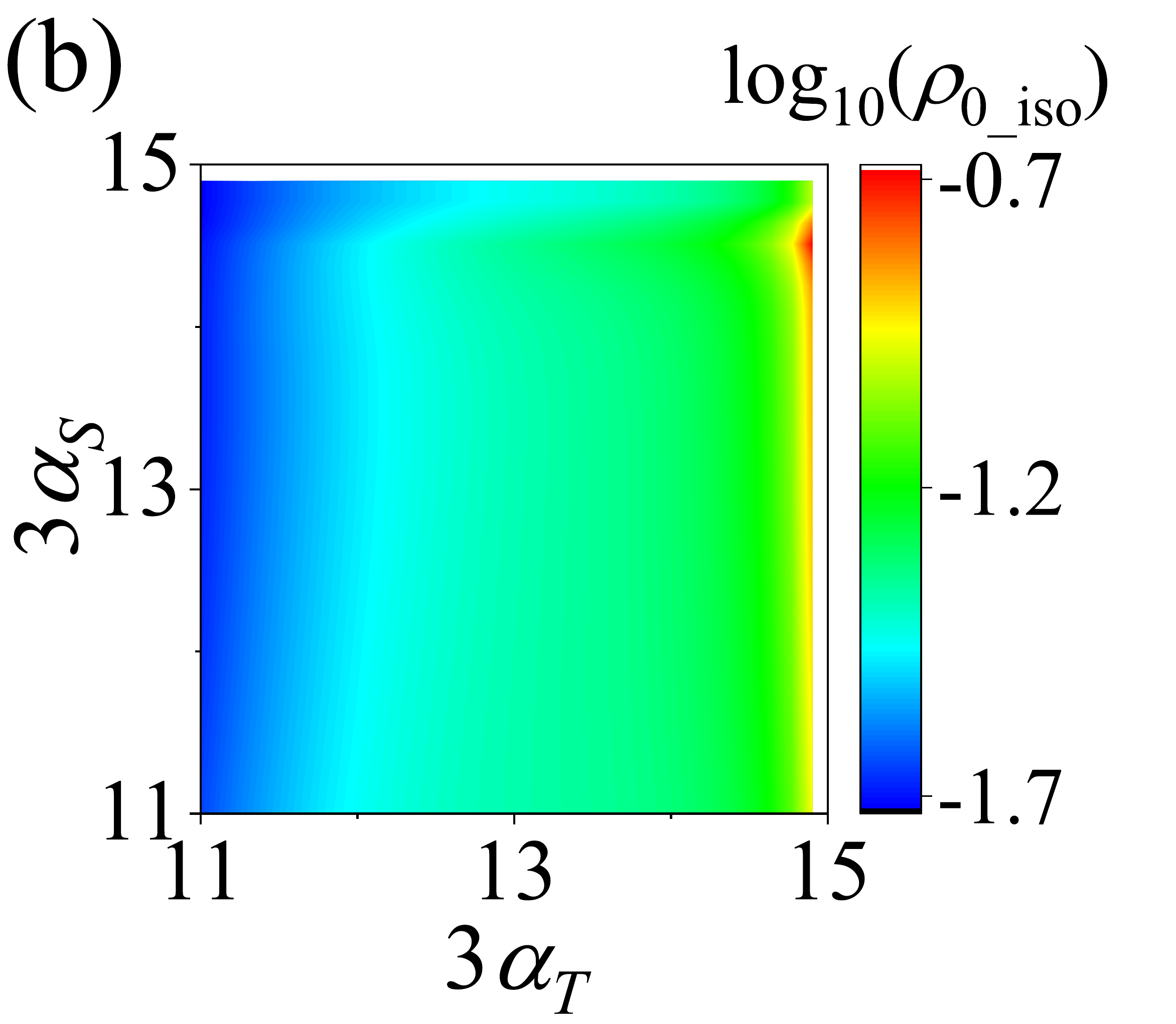}	
	}
	\subfigure[$C^2_S/C^2_T = 3.18\times10^{-1}$;]{
		\centering
		\includegraphics[width=0.26\linewidth]{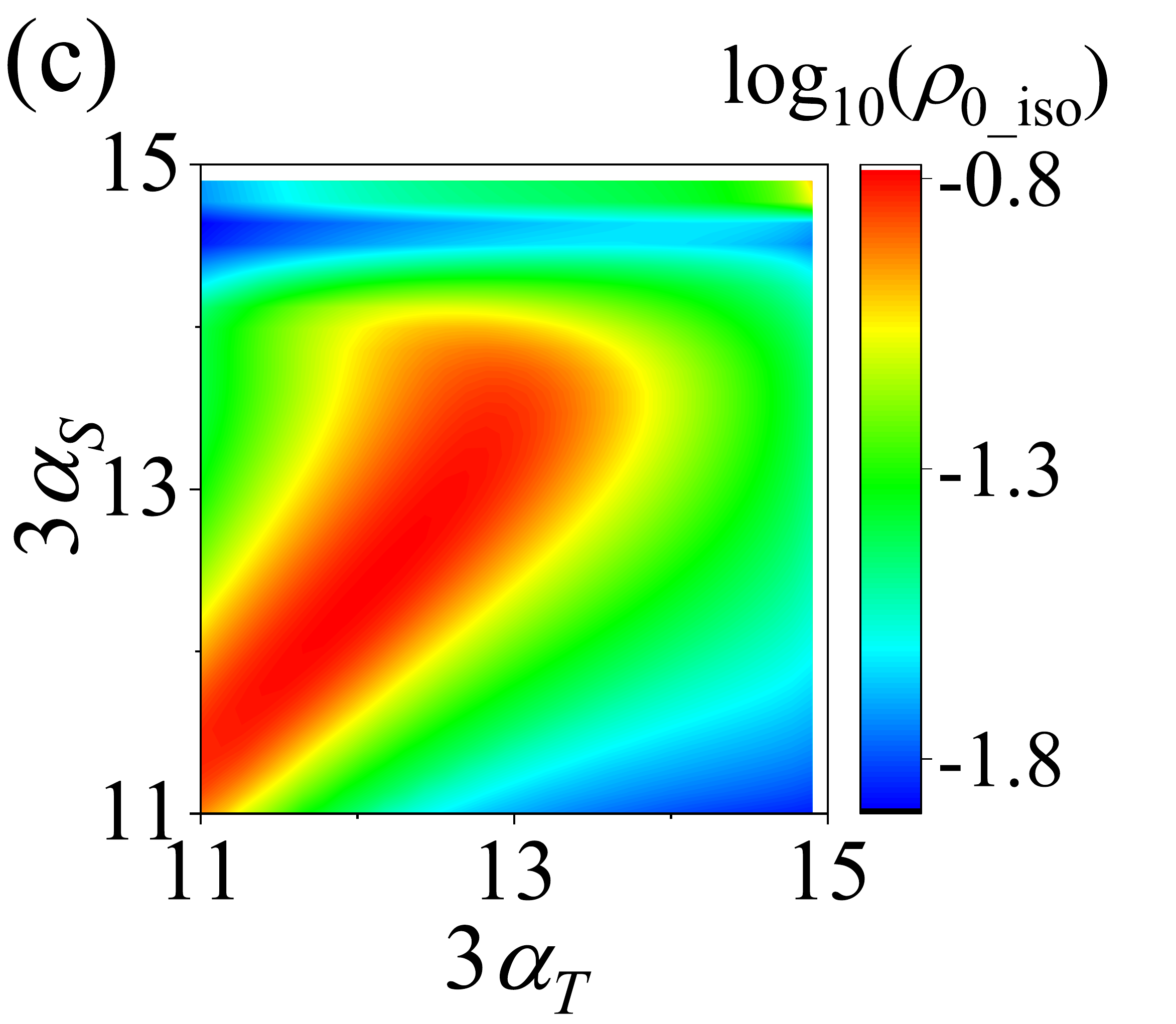}	
	}
	\\
	\subfigure[$C^2_S/C^2_T = 3.18\times10^{-5}$;]{
		\centering
		\includegraphics[width=0.26\linewidth]{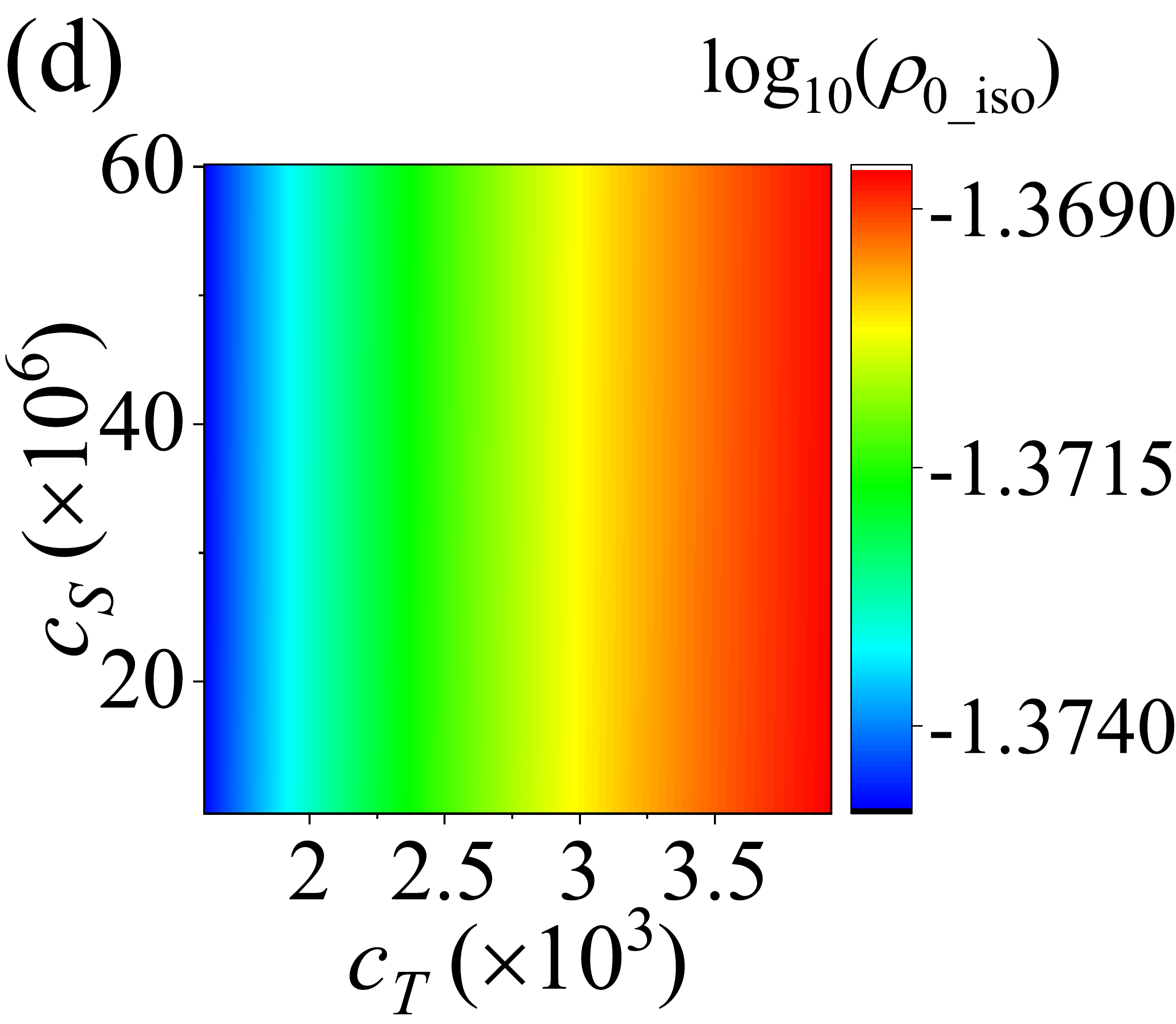}	
	}
	\subfigure[$C^2_S/C^2_T = 3.18\times10^{-3}$;]{
		\centering
		\includegraphics[width=0.26\linewidth]{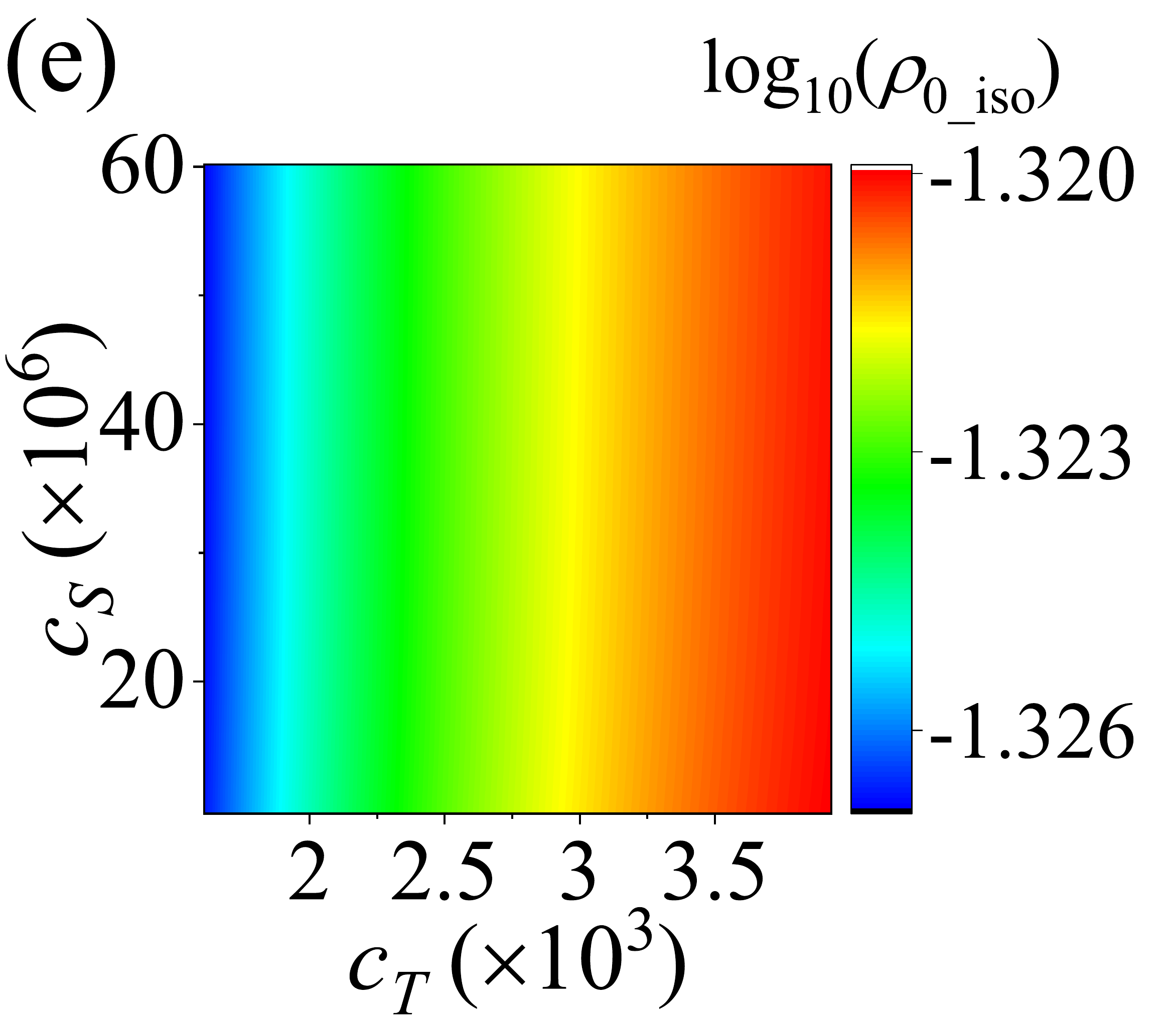}	
	}
	\subfigure[$C^2_S/C^2_T = 3.18\times10^{-1}$;]{
		\centering
		\includegraphics[width=0.26\linewidth]{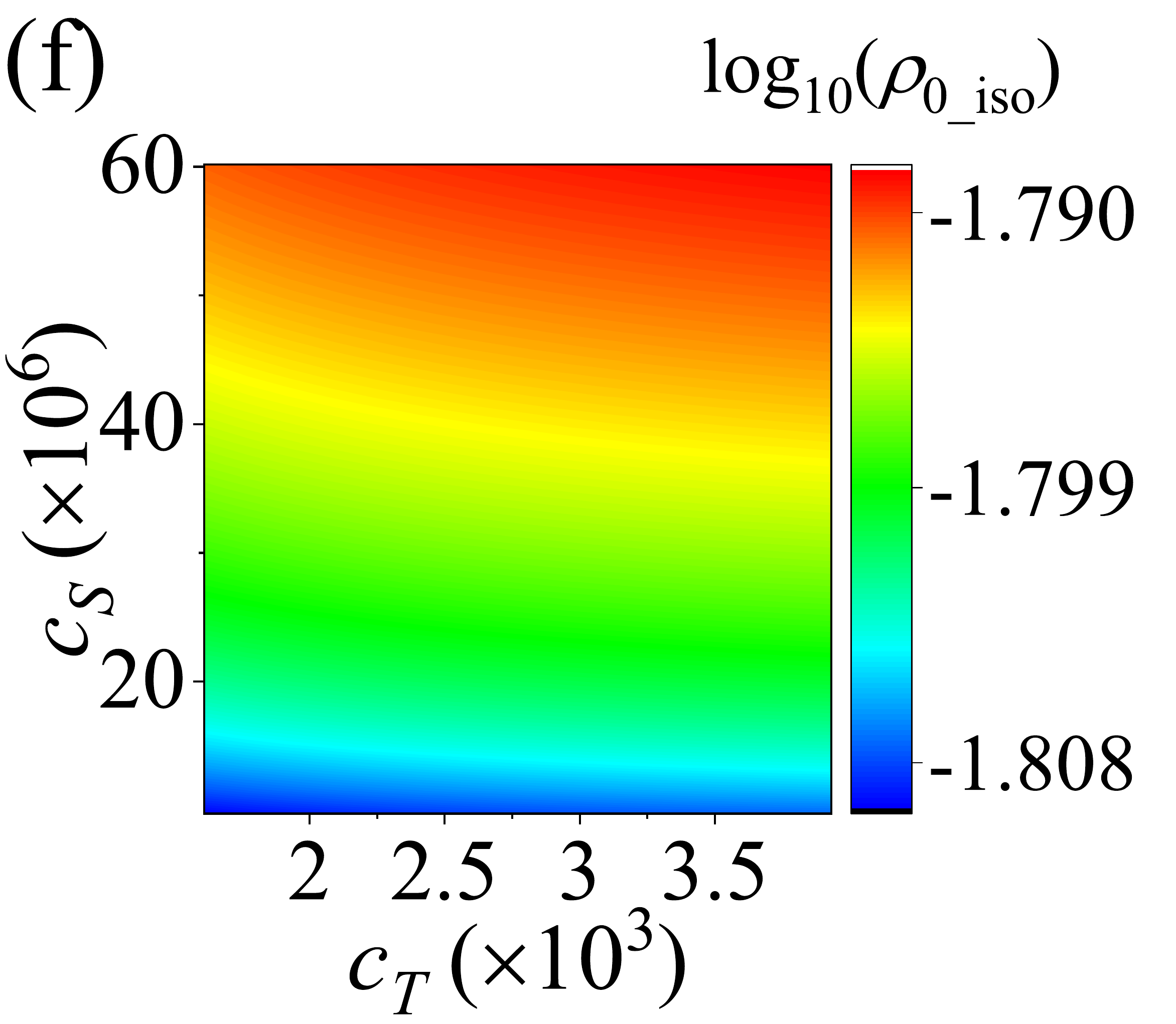}	
	}
	\caption{The distributions $\rho_{0\_{\rm{iso}}}\left(\alpha_{T}, \alpha_{S}\right)$ and $\rho_{0\_{\rm{iso}}}\left(c_T, c_S\right)$ with different values of $C^2_T/C^2_S$. All $C^2_S/C^2_T$ has the unit $\rm{ppt}^2\rm{deg}^{-2}\rm{m}^{\alpha_{T}-\alpha_{S}}$. $\alpha_T$, $\alpha_S$, $c_T$ and $c_S$ are dimensionless. Values of parameters are listed in Appendix II.}
	\label{fig:9}
\end{figure}

Figure \ref{fig:9} shows $\rho_{0\_{\rm{iso}}}\left(\alpha_{T}, \alpha_{S}\right)$ and $\rho_{0\_{\rm{iso}}}\left(c_T, c_S\right)$ with different ratios of $C^2_S$ to $C^2_T$. With the increase of $C^2_S/C^2_T$, the variation of $\rho_{0\_{\rm{iso}}}$ with $\alpha_{S}$ and $c_S$ becomes more pronounced.

A comprehensive analysis of Figs. \ref{fig:8} and \ref{fig:9} reveals that 
\begin{itemize}
	\item $\rho_{0\_{\rm{iso}}}$ substantially varies with $\alpha_{T}$ and $\alpha_{S}$ (can reach an order of difference magnitude).
	\item $\gamma_{ST}(\kappa\eta)$, as a function describing the correlation between temperature and salinity spectra, has an obvious effect on $\rho_{0\_{\rm{iso}}}$.
	\item As expected, the structure constant $C^2_T$ or/and $C^2_S$ describes the contribution of temperature or/and salinity fluctuation very well.
\end{itemize}

\section{Summary and conclusion}
The power spectrum of refractive-index fluctuations provides a rigorous physical description of the 2nd-order statistics of natural random media, hence, bearing utmost significance for environmental optics. A number of non-Kolmogorov models have been recently developed for `single-diffuser' turbulence, i.e.,  based on a single advected scalar, as is temperature in atmsopheric case. However, to our knowledge, there was no model for non-Kolmogorov spectrum describing optical turbulence with two or more advected scalars, i.e., `double-diffuser turbulence'. The major obstacle for developing such a power spectrum was due to the fact that the co-spectrum of two scalar spectra in the non-Kolmogorov case could not be directly obtained by analogy with a method used for Kolmogorov case in which the power laws of the two scalar spectra are equal.

In this paper, we have developed for the first time a non-Kolmogorov power spectrum of oceanic refractive-index fluctuations, being an example of a double-diffuser, by deriving the temperature spectrum, the salinity spectrum, and their co-spectrum, based on the Upper-Bound limitation and on the concept of spectral correlation. Our developed spectrum generally handles non-Kolmogorov turbulence with partially correlated temperature-salinity co-spectrum ($\alpha_i \in \left[11/3, 15/3\right)$ and $\gamma_{ST}(\kappa)\le1$) which is common for the stratified flow fields, but reduces to conventional, Kolmogorov spectrum, with fully correlated co-spectrum ($\alpha_i = 11/3$ and $\gamma_{ST}=1$). We have also provided the extension to anisotropic non-Kolmogorov turbulence case.

Besides, we have also illustarted how a non-Kolmogorov, isotropic and anisotropic oceanic turbulence affects the second-order statistics of a spherical wave. The numerical calculations have revealed that the turbulence's effect on a spherical wave substantially varies with the power law exponents ($\alpha_{T}$ and $\alpha_{S}$). Moreover, we have shown for the first time that the coherence radius scalar $\rho_{0\_\rm{iso}}$ takes on very different values for different settings of spectral correlation. This also indicates the usefulness of developing the oceanic non-Kolmogorov power spectrum with correlation factor $\gamma_{ST}$.

The numerical calculations in this manuscript focus on the coherence radius of a spherical wave. They are done for showing the significance of several parameters, like the non-Kolmogorov power law $\alpha_i$ and the correlation factor $\gamma_{ST}$ (they do change the coherence radius a lot). A further significant work is to obtain simple analytical formulae of coherence radius. This could be done with approximations and simplifying OTOPS. The readers can access the codes of numerical results at the link [https://github.com/jinry2017/Arxiv200902447].

On finishing we mention that so far no literature of oceanic turbulence has provided models for the correlation factor $\gamma_{ST}(\kappa)$ and other parameters such as $c_T$, $c_S$, $\alpha_{T}$ and $\alpha_{S}$. But like in the studies of atmospheric propagation, these parameters could be significant in characterizing oceanic optical turbulence, and any details about them are of importance for further experimental campaigns. Our model fills such a gap by providing a rather simple analytical model applicable in a variety of oceanic turbulence regimes. 

\section*{Appendix I. Ranges of parameters}

For brevity of numerical calculation, we set the ranges of parameters as follows. \textit{The ranges here are based on references, and some of them are obtained in Kolmogorov case. The real ranges could be beyond what we set.}
\subsection*{1. Constants}
As given in \cite{Nikishov_2000,muschinski2015direct,Hill1978}, $a = 0.072$, $\beta = 0.72$ and $Q = 2.73$.
\subsection*{2. The ranges of $\alpha_{T}$ and $\alpha_{S}$}
According to the experimental data in \cite{ichiye1972power} and the widely used range \cite{toselli2008}, non-Kolmogorov parameter $\alpha_i \in \left[11/3, 15/3\right)$.
\subsection*{3. The range of $C_S^2/C_T^2$}
According to Eq. (\ref{eq2}), 
\begin{equation}
{C_S^2}/{C_T^2}={\chi _S}/{\chi _T},
\label{eq56}
\end{equation}
where the dissipation rate $\chi_i$ of temperature and salinity are related through \cite{Elamassie:17,KOROTKOVA2020_1}
\begin{equation}
{\chi _S}/{\chi _T} = {d_r}{H^{-2}},
\label{eq57}
\end{equation}
with
\begin{equation}
{d_r} \approx \left\{ {\begin{array}{*{20}{l}}
	{\left| {H\theta_T {\theta_S ^{ - 1}}} \right| + {{\left| {H\theta_T {\theta_S ^{ - 1}}} \right|}^{0.5}}{{\left( {\left| {H\theta_T {\theta_S ^{ - 1}}} \right| - 1} \right)}^{0.5}},}&{\left| {H\theta_T {\theta_S ^{ - 1}}} \right| \ge 1,}\\
	{1.85\left| {H\theta_T {\theta_S ^{ - 1}}} \right| - 0.85,}&{0.5 \le \left| {H\theta_T {\theta_S ^{ - 1}}} \right| < 1,}\\
	{0.15\left| {H\theta_T {\theta_S ^{ - 1}}} \right|,}&{\left| {H\theta_T {\theta_S ^{ - 1}}} \right| < 0.5,}
	\end{array}} \right.
\label{eq58}
\end{equation}
where $d_r$ is the eddy diffusivity ratio, $\theta_T$ and $\theta_S$ are the thermal expansion coefficient and the saline contraction coefficient, respectively, and $H$ is the temperature-salinity gradient ratio defined by
\begin{equation}
H = \frac{{d\left\langle T \right\rangle /dz}}{{d\left\langle S \right\rangle /dz}}.
\label{eq59}
\end{equation}

\begin{figure*}
	\centering
	\includegraphics[width=0.9\linewidth]{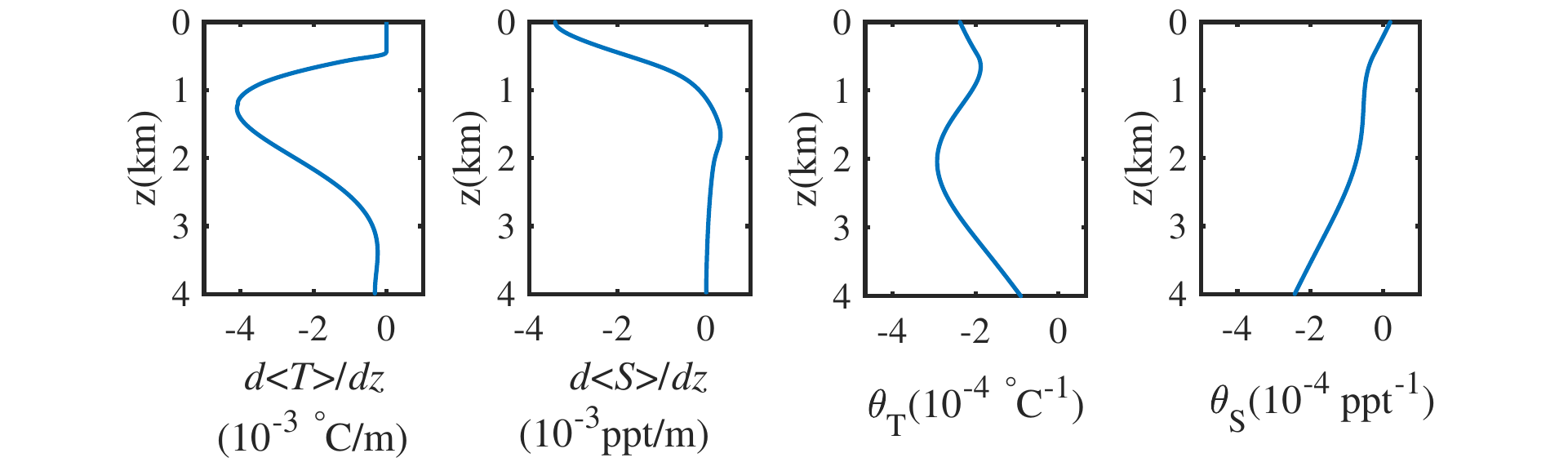}
	\caption{The distribution of temperature gradient $d\left\langle T \right\rangle/dz$, salinity gradient $d\left\langle S \right\rangle/dz$, thermal expansion coefficient $\theta_T$ and saline contraction coefficient $\theta_S$ varying with depth $z$ in Pacific.}
	\label{fig:10}
\end{figure*}

\begin{figure*}
	\centering
	\includegraphics[width=0.7\linewidth]{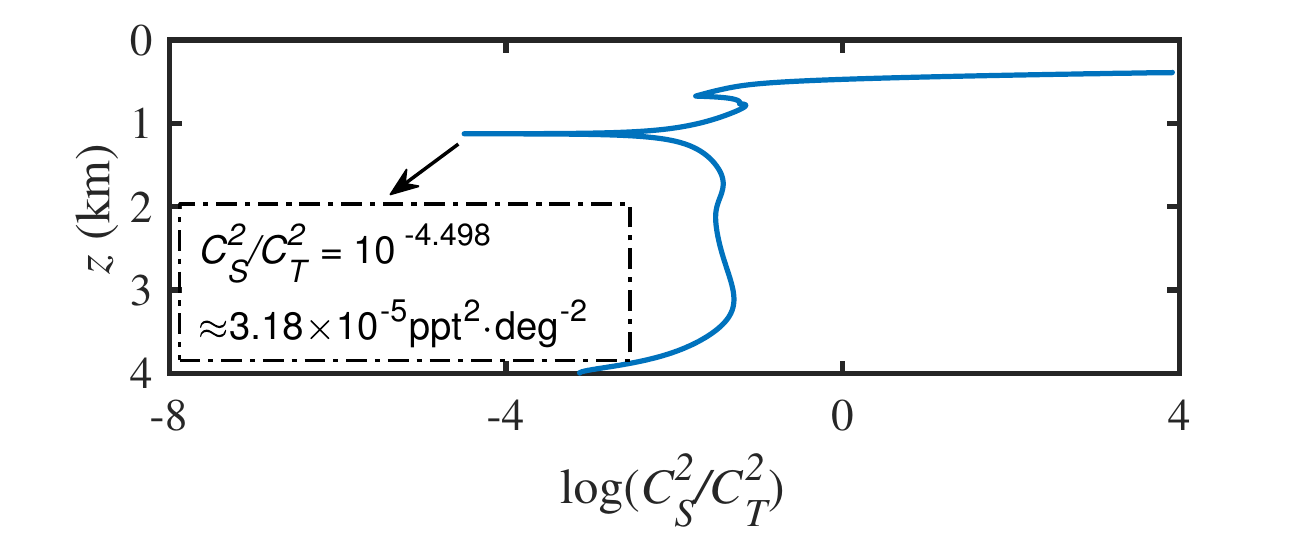}
	\caption{The distribution of $C^2_S/C^2_T$ [$\rm{ppt}^2\rm{deg}^{-2}\rm{m}^{\alpha_{T}-\alpha_{S}}$] varying with depth $z$ in Pacific.}
	\label{fig:11}
\end{figure*}

Combining Eqs.(\ref{eq56})-(\ref{eq59}), we have
\begin{equation}
\frac{{{C^2_S}}}{{{C^2_T}}} = \left\{ {\begin{array}{*{20}{l}}
	{\left| {{H^{ - 1}}\theta_T {\theta_S ^{ - 1}}} \right| + \left| {{H^{ - 1}}\theta_T {\theta_S ^{ - 1}}} \right|{{\left( {1 - \left| {{H^{ - 1}}\theta_S {\theta_T ^{ - 1}}} \right|} \right)}^{0.5}},}&{|H\theta_T {\theta_S ^{ - 1}}| \ge 1}\\
	{1.85\left| {{H^{ - 1}}\theta_T {\theta_S ^{ - 1}}} \right| - 0.85|H{|^{ - 2}},}&{0.5 \le |H\theta_T {\theta_S ^{ - 1}}| < 1}\\
	{0.15\left| {{H^{ - 1}}\theta_T {\theta_S ^{ - 1}}} \right|,}&{|H\theta_T {\theta_S ^{ - 1}}| < 0.5}
	\end{array}} \right.
\label{eq60}
\end{equation}
Using the data of $d\left\langle T \right\rangle /dz$, $d\left\langle S \right\rangle /dz$, $\theta_T$ and $\theta_S$ of mid latitude Pacific in winter \cite{Book03} (see also Fig. \ref{fig:10}), and based on Eq. (\ref{eq60}), we plot ${C_S^2}/{C_T^2}$ as a function of depth in Fig. \ref{fig:11}. It shows that
\begin{equation}
{C_S^2}/{C_T^2} \ge 3.18\times10^{-5}\rm{ppt}^2\cdot\rm{deg}^{-2}.
\label{eq61}
\end{equation}

For non-Kolmogorov cases, we assume
\begin{equation}
{C_S^2}/{C_T^2} \ge 3.18\times10^{-5}\rm{ppt}^2\cdot\rm{deg}^{-2}\cdot\rm{m}^{\alpha_{T}-\alpha_{S}}.
\label{eq62}
\end{equation}
\subsection*{4. The ranges of $c_S$ and $c_T$}
According to \cite{KOROTKOVA2020_1}, $Pr_T$ varies from 5.4 to 13.4, and $Pr_S$ varies from 350.0 to 2210.0. Using the relation in Eq.(\ref{eq6}) with constants $a = 0.072$ and $\beta = 0.72$, we have
\begin{equation}
c_T \in \left[1.61\times 10^{-3},3.99\times 10^{-3}\right]\quad \rm{and} \quad c_S \in \left[9.76\times 10^{-6},61.62\times 10^{-6}\right].
\label{eq63}
\end{equation}

\section*{Appendix II. The values of parameters in Figures}
Here we list the values of parameters in figures.
\begin{itemize}
	\item Figure \ref{fig:3}: $\alpha_{T}=\alpha_{S}=11/3$, $c_T = 2.6\times10^{-3}$, $c_S = 2.63\times10^{-5}$.
	\item Figure \ref{fig:5}: $c_T = 2.63\times10^{-3}$, $c_S = 2.55\times10^{-5}$, $C^2_T = 1.74\times10^{-4}\rm{deg}^2\rm{m}^{3-\alpha_{T}}$, $C^2_S =7.67\times10^{-6}\rm{ppt}^2\rm{m}^{3-\alpha_{S}}$, $\eta = 2.02\times10^{-4}\rm{m}$, $\lambda_0 = 532\rm{nm}$, $n'_T = -8.84\times10^{-5}\rm{deg}^{-1}1$, $n'_S = 1.87\times10^{-4}\rm{g}^{-1}1$, $L=15\rm{m}$, $L_0=30\rm{m}$.
	\item Figure \ref{fig:6}: $\alpha_{T} = 14/3$, $\alpha_{S} = 11/3$, $c_T = 2.63\times10^{-3}$, $c_S = 2.55\times10^{-5}$, $C^2_T = 1.74\times10^{-4}\rm{deg}^2\rm{m}^{3-\alpha_{T}}$, $C^2_S =7.67\times10^{-6}\rm{ppt}^2\rm{m}^{3-\alpha_{S}}$, $\eta = 2.02\times10^{-4}\rm{m}$, $\lambda_0 = 532\rm{nm}$, $n'_T = -8.84\times10^{-5}\rm{deg}^{-1}1$, $n'_S = 1.87\times10^{-4}\rm{g}^{-1}1$, $L=15\rm{m}$, $L_0=30\rm{m}$.
	\item Figure \ref{fig:7}: same as those values in Fig. \ref{fig:6}.
	\item Figure \ref{fig:8}: $C^2_T = 1.74\times10^{-4} \rm{deg}^2\rm{m}^{3-\alpha_{T}}$, $C^2_S = 7.67\times10^{-6} \rm{ppt}^2{m}^{3-\alpha_{S}}$, $\lambda_0 = 532\rm{nm}$, $n'_T = -8.84\times10^{-5}\rm{deg}^{-1}1$, $n'_S = 1.87\times10^{-4}\rm{g}^{-1}1$, , $L=15\rm{m}$, $L_0=30\rm{m}$, and $\eta = 2.02\times10^{-4}\rm{m}$. (a)-(c) are plotted with $(c_T,c_S) = (2.63\times10^{-3}, 2.55\times10^{-5})$, and (d)-(e) are plotted with $(\alpha_{T},\alpha_{S}) = (14/3,11/3)$.
	\item Figure \ref{fig:9}: $C^2_T = 1.74\times10^{-4} \rm{deg}^2\rm{m}^{3-\alpha_{T}}$, $\lambda_0 = 532\rm{nm}$, $n'_T = -8.84\times10^{-5}$, $n'_S = 1.87\times10^{-4}$, $\eta = 2.02\times10^{-4}$, $L=15\rm{m}$, $L_0=30\rm{m}$, $\gamma$ is given by Eq. (\ref{eq31}) with $p=3$. (a)-(c) are plotted with $(c_T,c_S) = (2.63\times10^{-3}, 2.55\times10^{-5})$, and (d)-(e) are plotted with $(\alpha_{T},\alpha_{S}) = (14/3,11/3)$.
\end{itemize}

\section*{Appendix III. Terminologies}
Here we list a brief explanation about some terminology in this manuscript.

\begin{itemize}
	\item 	\textbf{Coherence radius vector} (CRV) and \textbf{coherence radius scalar} (CRS):\\
	According to Section 4.1, the WSF $D_{sp}$ in anisotropic turbulence could be also anisotropic. Hence, the coherence radius $\left|\bm{\rho_0}\right|$ in $D_{sp}(\bm{\rho_0})=2$ could vary with the orientation of $\bm{\rho_0}$. For brevity in discussion, we define $\bm{\rho_0}$ as CRV, and define a scalar --- CRS --- in Eq. (\ref{eq54}). The CRS equals coherence radius if $\mu = 1$ or in vertical channels.
	\item \textbf{Hill's model 1} (H1) and \textbf{Hill's model 4} (H4):\\
	As widely accepted, the power spectrum of scalar fluctuations has two or three intervals \cite{sreenivasan2019turbulent}. For the turbulence with large Pr or Sc, there are three intervals: inertial-convective, viscous-convective and viscous-diffusive intervals.  For the turbulence with small Pr or Sc, there are two intervals: inertial and diffusive intervals. Hill's models provide continuous transition between different intervals. Hill's model 1 is mathematically analytic but not as precise as Hill's model 4, and Hill's model 4 is a non-linear differential equation that does not have a closed-form solution. By numerical fitting, some approximate models for ocean \cite{Yao_19,Yi_18} and atmosphere \cite{Andrews_92} have been proposed based on Hill's model 4.
	\item \textbf{H1-based} and \textbf{H4-based}:\\
	They refer to the models based on Hill's model 1 and 4, respectively.
	\item \textbf{Upper-bound limitation}:\\
	As proved in the Section 5.2.5 of \cite{Bernard1975}, the co-spectrum $\phi_{ab}$ of scalars $a$ and $b$ are limited by
	\begin{equation}
	\left|\phi_{ab}\right|^2 \le \phi_{a}\phi_{b},
	\label{eq65}
	\end{equation}
	where $\phi_a$ and $\phi_b$ are the spectra of $a$ and $b$, respectively.
	\item \textbf{spectral correlation, fully correlated, partially correlated} and \textbf{uncorrelated}:\\
	The `Correlation' in this manuscript refers to the correlation between temperature fluctuations and salinity fluctuations. The spectral correlation factor is defined as
	\begin{equation}
	\gamma_{ST} = \left[\frac{\left|\Phi_{TS}\right|^2}{\Phi_{T}\Phi_{S}}\right]^{1/2},
	\label{eq66}
	\end{equation}
	where $\Phi_T$ and $\Phi_S$ are the 3-D spectra of temperature and salinity, respectively.
	`fully correlated' and `full correlation' refer to the cases of $\gamma_{ST} = 1$; `partially correlated' and `partial correlation' refer to the cases of $\gamma_{ST} \textless 1$; `uncorrelated' and `non-correlation' refer to the cases of $\gamma_{ST} = 0$.
\end{itemize}

\section*{Disclosures}
The authors declare no conflicts of interest.


\begin{thebibliography}{10}
\bibitem{footnote3}
	Generally, salinity Schmidt number is so larger than temperature
		Prandtl number that the inertial-convective range of salinity is wider than
		that of temperature. Hence, the salinity fluctuation is in its
		inertial-convective range if $\kappa$ locates in the inertial-convective
		range of temperature spectrum.
\bibitem{footnote2}
	The concept of spectral correlation has not been introduced into the
		H4-based temperature-salinity co-spectrum \cite{Yao_19}. Here we calculate
		the $\gamma_{ST}$ of traditional co-spectrum by calculating
		$\Phi_{TS}/\left(\Phi_T\Phi_S\right)^{1/2}$.
\end{thebibliography}

\end{document}